\documentclass[10pt]{scrartcl}
\pdfoutput=1

\usepackage[utf8]{inputenc}

\usepackage{amsmath,amsfonts}
\usepackage{booktabs}
\usepackage{cite}
\usepackage{graphicx, subfigure, ulem}
\usepackage{xcolor}
\usepackage{siunitx}
\usepackage{cite}
\usepackage{soul}
\usepackage{wasysym}              
\usepackage{hyphenat}             

\usepackage[margin=0.8cm, font=small, labelfont=bf]{caption}
\DeclareCaptionSubType*[arabic]{figure}
\DeclareCaptionSubType*[arabic]{table}
\usepackage{etoolbox}

\let\theparentequation\theequation
\patchcmd{\theparentequation}{equation}{parentequation}{}{}

\renewenvironment{subequations}[1][]{
  \refstepcounter{equation}%
  \setcounter{parentequation}{\value{equation}}
  \setcounter{equation}{0}
  \def\theequation{\theparentequation\alph{equation}}%
  \let\parentlabel\label
  \ifx\\#1\\\relax\else\label{#1}\fi
  \ignorespaces
}{%
  \setcounter{equation}{\value{parentequation}}
  \ignorespacesafterend
}

\newcommand*{\nextParentEquation}[1][]{
  \refstepcounter{parentequation}
  \setcounter{equation}{0}
  \ifx\\#1\\\relax\else\parentlabel{#1}\fi
}

\makeatletter
\renewcommand\@makefntext[1]{\leftskip=.5em\hskip-.5em\@makefnmark#1}
\makeatother

\usepackage{jheppub}

\let\theparentequation\theequation
\patchcmd{\theparentequation}{equation}{parentequation}{}{}

\newcommand{\IE}{\textit{i.\,e.}}
\newcommand{\EG}{\textit{e.\,g.}}
\newcommand{\I}{{i\mkern1mu}}
\newcommand{\E}{\mathrm{e}}
\newcommand{\Real}[1]{\Re\hspace{-1pt}\mathfrak{e}{\left[#1\right]}}

\newcommand{\Imag}[1]{\Im\hspace{-1pt}\mathfrak{m}{\left[#1\right]}}

\newcommand{\mueff}{\ensuremath{\mu_{\text{eff}}}}
\newcommand\MHp{\ensuremath{M_{H^\pm}}}
\newcommand\tb{\ensuremath{\tan{\beta}}}
\newcommand{\MW}{M_W}
\newcommand{\MZ}{M_Z}
\newcommand{\cp}{\ensuremath{{\cal CP}}}

\newcommand{\SUL}{\ensuremath{SU(2)_\text{L}}}
\newcommand\sw{\ensuremath{s_\mathrm{w}}}
\newcommand\cw{\ensuremath{c_\mathrm{w}}}
\newcommand{\FH}{\texttt{FeynHiggs}}
\newcommand{\NC}{\texttt{NMSSMCALC}}
\newcommand{\NCs}{\texttt{NC}}
\newcommand{\FA}{\texttt{FeynArts}}

\newcommand{\HS}{\texttt{HiggsSignals}}
\newcommand{\HB}{\texttt{HiggsBounds}}
\newcommand{\Zmix}{\ensuremath{\mathbf{Z}^{\mbox{\tiny mix}}}}

\newcommand{\DRbar}{\ensuremath{\overline{\mathrm{DR}}}}

\newcommand{\MHexp}{\SI{125}{\giga\electronvolt}}
\newcommand{\PrL}{\ensuremath{P_{\textrm{L}}}}
\newcommand{\PrR}{\ensuremath{P_{\textrm{R}}}}

\usepackage{setspace}

\BeforeTOCHead[toc]{{\pdfbookmark[1]{\contentsname}{toc}}}

\let\OLDthebibliography\thebibliography
\renewcommand\thebibliography[1]{
  \OLDthebibliography{#1}
  \setlength{\parskip}{1pt}
  \setlength{\itemsep}{0pt plus 0.3ex}
}

\begin{document}


\thispagestyle{empty}

\def\thefootnote{\fnsymbol{footnote}}

\begin{flushright}
DESY--17--067\\
IFT--UAM/CSIC--17--042
\end{flushright}

\vspace{2cm}

\begin{center}

{\large\textsc{\textbf{On-Shell neutral Higgs bosons in}}}

\vspace{0.4cm}

{\large\textsc{\textbf{the NMSSM with complex parameters}}}

\vspace{1cm}

Florian Domingo$^{1,2}$\footnote{email: florian.domingo@csic.es},
Peter Drechsel$^3$\footnote{email: peter.drechsel@desy.de},
and
Sebastian Pa{\ss}ehr$^3$\footnote{email: sebastian.passehr@desy.de}

\vspace*{.7cm}

\textsl{
$^1$Instituto de Física Teórica (UAM/CSIC), Universidad Autónoma de Madrid,
Cantoblanco, E-28049 Madrid, Spain
}

\medskip
\textsl{
$^2$Instituto de Física de Cantabria (CSIC-UC), E-39005 Santander, Spain
}


\medskip
\textsl{
$^3$Deutsches Elektronensynchrotron DESY, Notkestraße 85, D--22607 Hamburg, Germany
}

\end{center}

\vspace*{2cm}

\begin{abstract}
\noindent
The Next-to-Minimal  Supersymmetric Standard model (NMSSM)  appears as
an   interesting    candidate   for   the   interpretation    of   the
Higgs-measurement at the LHC and as a rich framework embedding physics
beyond  the Standard  Model. We  consider the  renormalization of  the
Higgs sector of this model in its \cp-violating version, and propose a
renormalization scheme  for the calculation of  on-shell Higgs masses.
Moreover,  the   connection  between  the  physical   states  and  the
tree-level ones is no longer trivial  at the radiative level: a proper
description of  the corresponding transition thus  proves necessary in
order to  calculate Higgs production  and decays at a  consistent loop
order.  After discussing these formal  aspects, we compare the results
of our  mass calculation  to the  output of  existing tools.   We also
study the relevance  of the on-shell transition-matrix  in the example
of the $h_i\to\tau^+\tau^-$ width. We find deviations between our full
prescription and popular approximations that can exceed $10\%$.
\end{abstract}

\def\thefootnote{\arabic{footnote}}
\setcounter{page}{0}
\setcounter{footnote}{0}

\newpage
\tableofcontents

\section{\label{sec:intro}Introduction}

Since  the discovery  of  a  Higgs-like particle  with  a mass  around
\MHexp\   by   the   ATLAS  and   CMS   experiments~\cite{Aad:2012tfa,
  Chatrchyan:2012ufa} at  CERN, a lot  of effort has been  invested to
reveal its nature as the particle responsible for electroweak symmetry
breaking.   While within  the present  experimental uncertainties  the
properties of the  observed state are compatible  with the predictions
of  the  Standard  Model  (SM)~\cite{Khachatryan:2016vau}  many  other
interpretations are possible  as well, in particular as  a Higgs boson
of an extended Higgs sector.

One of the prime candidates for physics beyond the SM is softly-broken
supersymmetry (SUSY), which doubles the particle degrees of freedom by
predicting  two  scalar partners  for  each  SM  fermion, as  well  as
fermionic      partners     for      all     bosons---for      reviews
see~\cite{Nilles:1983ge,Haber:1984rc}.        The      Next-to-Minimal
Supersymmetric            Standard            Model            (NMSSM)
\cite{Ellwanger:2009dp,Maniatis:2009re} is  a well-motivated extension
of  the SM.   In particular  it provides  a solution  for the  ``$\mu$
problem''~\cite{Kim:1983dt}  of  the Minimal  Supersymmetric  Standard
Model (MSSM), by naturally relating the $\mu$ parameter to a dynamical
scale of the Higgs potential~\cite{Ellis:1988er, Miller:2003ay}.

In contrast to the single Higgs doublet in the SM, the Higgs sector of
the NMSSM  contains two Higgs doublets  (like the MSSM) and  one Higgs
singlet.  After  electroweak symmetry  breaking the  physical spectrum
consists of five neutral Higgs bosons,  $h_i$ ($i \in [1,5]$), and the
charged Higgs boson pair $H^\pm$.  Ever since the Higgs discovery, the
possibility to  interpret this  signal in terms  of an  NMSSM (mostly)
\cp-even Higgs boson has been emphasized in Refs.~\cite{Gunion:2012zd,
  Ellwanger:2012ke,   Gunion:2012gc,    Kowalska:2012gs,   Bae:2012am,
  Beskidt:2013gia, Moretti:2013lya,  Moretti:2015bua, Domingo:2015eea,
  Carena:2015moc}.   In particular,  it has  been argued  that such  a
solution  came   with  improved  naturalness  compared   to  the  MSSM
interpretation~\cite{Hall:2011aa,    Arvanitaki:2011ck,   Kang:2012sy,
  Cao:2012yn,      Kyae:2012rv,      Jeong:2012ma,      Agashe:2012zq,
  Gherghetta:2012gb, Barbieri:2013hxa}.  Moreover,  several works have
pointed out the possibility to accommodate deviations from a strict SM
behavior  in  the  diphoton  rate, in  Higgs-pair  production  or  in
associated        production~\cite{Ellwanger:2011aa,       Cao:2012fz,
  Benbrik:2012rm,      Heng:2012at,     Choi:2012he,      King:2012tr,
  Vasquez:2012hn,    Jia-Wei:2013eea,    Munir:2013wka,    Cao:2013si,
  Nhung:2013lpa, Christensen:2013dra,  Ellwanger:2013ova, Han:2013sga,
  Cao:2014kya, Wu:2015nba,  Badziak:2016tzl,Das:2017tob}.  Admittedly,
the viability of the extended NMSSM Higgs sector would be comforted by
the detection of additional Higgs  states. To this end, several search
channels  have  been suggested,  especially  for  states lighter  than
$125$\,GeV~\cite{Belanger:2012tt,  Almarashi:2012ri,  Rathsman:2012dp,
  Cerdeno:2013cz,   Barbieri:2013nka,  Badziak:2013bda,   Kang:2013rj,
  Cao:2013gba,    King:2014xwa,     Bomark:2014gya,    Bomark:2015fga,
  Buttazzo:2015bka, Guchait:2015owa, Ellwanger:2015uaz, Conte:2016zjp,
  Guchait:2016pes, Das:2016eob, Cao:2016uwt}.   Another feature of the
NMSSM  phenomenology is  the extended  neutralino sector,  due to  the
singlino.

In contrast  to the situation  in the MSSM, \cp-violation  can already
occur     at     the     tree-level     in     the     NMSSM     Higgs
sector~\cite{Garisto:1993ms,        Matsuda:1995ta,       Haba:1995aw,
  Asatrian:1996wn,     Haba:1996bg,     Ham:1999zs,     Branco:2000dq,
  Davies:2001uv,      Ham:2001wt,     Ham:2001kf,      Hugonie:2003yu,
  Miller:2003ay, Ham:2003jf,  Funakubo:2004ka, Ham:2007mt, Ham:2008cg,
  Cheung:2010ba,  Moretti:2013lya,  Munir:2013dya}.  While  low-energy
observables      place      limits     on      such      \cp-violating
scenarios~\cite{King:2015oxa},   especially    on   MSSM-like   phases
\cite{Arbey:2014msa},  \cp-violation  beyond  the   SM  appears  as  a
well-motivated        requirement        for       a        successful
baryogenesis~\cite{Sakharov:1967dj}. Correspondingly, several computer
tools have been proposed in the past few years to promote the study of
the     \cp-violating     NMSSM:     \verb|SPHENO|~\cite{Porod:2003um,
  Porod:2011nf,        Goodsell:2014bna,        spheno-www}        and
\verb|FlexibleSUSY|~\cite{Athron:2014yba,    FlexibleSUSY-www}---which
employ
\verb|SARAH|~\cite{Staub:2009bi,Staub:2010jh,Staub:2012pb,Staub:2013tta}
in  order to  produce their  modelfiles; \verb|FlexibleSUSY|  contains
components         from         \verb|SoftSUSY|~\cite{Allanach:2001kg,
  Allanach:2013kza}  and only  the \cp-conserving  case is  explicitly
mentioned for both---as well as \verb|NMSSMCALC|~\cite{Baglio:2013iia,
  NMSSMCALC-www}     and     \verb|NMSSMTools|~\cite{Ellwanger:2004xm,
  Ellwanger:2005dv, Domingo:2015qaa, NMSSMTOOLS-www}.

In this  work, we  specialize in the  $Z_3$-conserving version  of the
NMSSM, characterized  by a  scale-invariant superpotential.   The main
effort of our  project consists in analyzing  radiative corrections in
the Higgs sector of the \cp-violating NMSSM. To serve this purpose, we
elaborated a  \verb|FeynArts|~\cite{Kublbeck:1990xc,Hahn:2000kx} model
file and a set of  \texttt{Mathematica} routines for the evaluation of
the  Higgs  masses  and  wave-function normalization  matrix  at  full
one-loop order and beyond. These should  serve as a basis for a future
inclusion      of     the      \cp-violating     NMSSM      in     the
\verb|FeynHiggs|~\cite{Heinemeyer:1998np,           Heinemeyer:1998yj,
  Degrassi:2002fi,  Frank:2006yh,Hahn:2010te,   Bahl:2016brp,  FH-www}
package---originally designed for precise  calculations of the masses,
decays, and other properties of the Higgs bosons in the \cp-conserving
or -violating MSSM.  A first step  in this direction is represented by
Ref.~\cite{Drechsel:2016jdg}, centering  on the  \cp-conserving NMSSM.
In the current  paper, we expand this project further.   We follow the
general    methodology    of    \verb|FeynHiggs|,   relying    on    a
Feynman-diagrammatic  calculation  of   radiative  corrections,  which
employs            \verb|FeynArts|~\cite{Kublbeck:1990xc,Hahn:2000kx},
\verb|FormCalc|~\cite{Hahn:1998yk}                                 and
\verb|LoopTools|~\cite{Hahn:1998yk}.    Our   chosen   renormalization
scheme  differs  somewhat  from  earlier  proposals~\cite{Graf:2012hh,
  Goodsell:2014bna,   Drechsel:2016jdg}.    In  particular,   in   our
renormalization  scheme, the  electromagnetic coupling  $e$---which is
related  to the  fine-structure  constant  $\alpha =  e^2/(4\pi)$---is
defined in terms of the Fermi constant $G_F$ measured in muon decays.

In section~\ref{sec:theory}, we shall introduce relevant notations and
describe the renormalization procedure underpinning our model file for
the  \cp-violating  NMSSM.   In  this section  we  also  describe  our
implementation  of higher-order  corrections in  the Higgs  sector.  A
numerical     evaluation     of     our     results     follows     in
section~\ref{sec:numerics}, where we will  validate our calculation by
a comparison with  public codes. We will also insist  on the relevance
of the field-renormalization matrix for a consistent evaluation of the
Higgs  decays at  the one-loop  level,  before a  short conclusion  in
section~\ref{sec:conslusion}.

\section{\label{sec:theory}\boldmath Higgs masses and mixing in the \cp-violating NMSSM}

After a few general remarks  concerning our notations and conventions,
we  present  the renormalization  conditions  that  we employ  in  our
calculation.  There, we focus on effects  beyond the MSSM in the Higgs
and higgsino sectors, since we otherwise align with the conventions of
\texttt{FeynHiggs}, described in~\cite{Frank:2006yh}.  Then we discuss
how to formally  extract the loop-corrected Higgs masses  and the wave
function normalization factors.

\subsection{Conventions and relations at the tree level}

In  the following,  we consider  the $Z_3$-conserving  version of  the
NMSSM  and neglect  flavor-mixing.  The  superpotential of  the NMSSM
(showing only one generation of fermions/sfermions) reads
\begin{align}
\label{eq:superpot}
  W = 
  Y_u\,\hat{u} \left(\hat{H}_2\cdot \hat{Q} \right)
  - Y_d\,\hat{d} \left(\hat{H}_1\cdot \hat{Q} \right)
  - Y_e\,\hat{e} \left(\hat{H}_1\cdot \hat{L} \right)
  + \lambda\,\hat{S} \left(\hat{H}_2 \cdot \hat{H}_1 \right) 
  + \frac{1}{3}\,\kappa\,\hat{S}^3,
\end{align}
\noindent
where   $\hat{Q}$,   $\hat{u}$,   $\hat{d}$,   $\hat{L}$,   $\hat{e}$,
$\hat{H}_1$, $\hat{H}_2$, $\hat{S}$ denote the quark, lepton and Higgs
superfields.  The dot $\cdot$ stands for the $\SUL$-invariant product.
The  Yukawa couplings  in  Eq.~\eqref{eq:superpot} can  be complex  in
general.  However, their  phases can be absorbed in  a redefinition of
the quark and  lepton superfields.  We may write the  scalar fields in
$\hat{H}_1$, $\hat{H}_2$  and $\hat{S}$  explicitly in terms  of their
(real and positive) vacuum expectation values (vevs), $v_1$, $v_2$ and
$v_s$, respectively, as  well as their \cp-even,  \cp-odd, and charged
components, $\phi_i$, $\chi_i$, and $\phi^\pm_i$,
\begin{align}
\label{eq:vev}
  \mathcal{H}_1 &= 
  \E^{\I\,\xi_1}
  \begin{pmatrix} 
    v_1 + \frac{\left( \phi_1 + \I\chi_1\right)}{\sqrt{2}}
    \\
    \phi_1^- 
  \end{pmatrix},
  &
  \mathcal{H}_2 &=
  \E^{\I\,\xi_2}
  \begin{pmatrix}
    \phi_2^+
    \\
    v_2 + \frac{\left( \phi_2 + \I\chi_2\right)}{\sqrt{2}}
  \end{pmatrix},
  &
  \mathcal{S} &=
  \E^{\I\,\xi_s} \left[v_s + \tfrac{\left( \phi_s+i\chi_s\right)}{\sqrt{2}}\right].
\end{align}
\noindent
Here $\xi_1$,  $\xi_2$ and  $\xi_s$ are  the phases  of the  two Higgs
doublets and  the Higgs  singlet, respectively.   It is  convenient to
define the  ratio $\tb = v_2/v_1$,  the geometric mean of  the doublet
vevs $v  = \sqrt{v_1^2 + v_2^2}$,  as well as  the sum $\xi =  \xi_1 +
\xi_2$ of the doublet phases.  Since $\hat{S}$ transforms as a singlet
under  the  SM-gauge  transformations,  the $D$-terms  of  the  scalar
potential are unchanged with respect to  the MSSM.  On the other hand,
as compared to the  MSSM, additional dimensionless, complex parameters
$\lambda=\lvert\lambda\rvert\,\E^{\I\,\phi_\lambda}$               and
$\kappa=\lvert\kappa\rvert\,\E^{\I\,\phi_\kappa}$  appear   while  the
complex $\mu$-term is absent.  The  latter is dynamically generated as
an effective $\mu$-term when the singlet field takes its vev,
\begin{align}
  \mueff = \lvert\mueff\rvert\,
  \E^{\I\,\phi_\mu} = \lvert\lambda\rvert\,
  v_s\,\E^{\I\left(\phi_\lambda + \xi_s\right)}.
\end{align}
\noindent
In  the  NMSSM  the  phases  $\xi$ and  $\xi_s$  only  appear  in  the
combinations $\phi_{\lambda}+\xi_s+\xi$  and $\phi_{\kappa}+3\,\xi_s$,
so that they could be  absorbed in a re-definition of $\phi_{\lambda}$
and $\phi_{\kappa}$.  Nevertheless, we will keep the dependence on all
phases of  the Higgs  sector explicitly,  in order  to allow  for more
flexibility on the choice of input.

Soft  SUSY-breaking  in  the  NMSSM is  parametrized  by  the  complex
trilinear    soft-breaking    parameters     $A_\lambda    =    \lvert
A_\lambda\rvert\,\E^{\I\,\phi_{A_\lambda}}$,     $A_\kappa=     \lvert
A_\kappa\rvert\,\E^{\I\,\phi_{A_\kappa}}$,      $A_u     =      \lvert
A_u\rvert\,\E^{\I\,\phi_{A_u}}$,         $A_d         =         \lvert
A_d\rvert\,\E^{\I\,\phi_{A_d}}$,      and      $A_e      =      \lvert
A_e\rvert\,\E^{\I\,\phi_{A_e}}$,  as well  as  the real  soft-breaking
mass  terms  $m_{1,2}^2$  and  $m_S^2$   for  the  Higgs  fields,  and
$m^2_{\tilde{Q}}$,        $m^2_{\tilde{U}}$,        $m^2_{\tilde{D}}$,
$m^2_{\tilde{L}}$ and $m^2_{\tilde{E}}$ for the sfermions,
\begin{align}
\label{eq:soft}
  \begin{split}
    \mathcal{L}_{\text{soft}} &= 
    - m_1^2\,\lvert\mathcal{H}_{1}\rvert^2
    - m_2^2\,\lvert\mathcal{H}_{2}\rvert^2
    - m_S^2\,\lvert\mathcal{S}\rvert^2
    - \left[
      \lambda\, A_\lambda\, \mathcal{S} \left(\mathcal{H}_2 \cdot \mathcal{H}_1\right)
      +
      \tfrac{1}{3}\,\kappa\, A_\kappa\, \mathcal{S}^3 + \text{h.c.}
    \right]\\
    &\quad
    - m_{\tilde{Q}}^2\,\lvert\tilde{Q}\rvert^2
    - m_{\tilde{U}}^2\,\lvert\tilde{u}\rvert^2
    - m_{\tilde{D}}^2\,\lvert\tilde{d}\rvert^2
    - m_{\tilde{L}}^2\,\lvert\tilde{L}\rvert^2
    - m_{\tilde{E}}^2\,\lvert\tilde{e}\rvert^2\\
    &\quad
    - \left[
      -Y_u\, A_u\, \tilde{u} \left(\mathcal{H}_2 \cdot \tilde{Q}\right)+
      Y_d\, A_d\, \tilde{d} \left(\mathcal{H}_1 \cdot \tilde{Q}\right)+
      Y_e\, A_e\, \tilde{e} \left(\mathcal{H}_1 \cdot \tilde{L}\right)
      + \text{h.c.}
    \right]\,.
  \end{split}
\end{align}
\noindent
Expanding the  Higgs potential  in terms of  the charged  Higgs fields
$(\phi_1^+,\phi_2^+)  =  (\phi_1^-,\phi_2^-)^*$,   and  neutral  Higgs
fields  $\left(\phi,\chi\right)=\left(\phi_1, \phi_2,  \phi_s, \chi_1,
\chi_2, \chi_s\right)$ yields
\begin{align}
  \label{eq:HiggsPot}
  V_{\text{H}} =
  -\mathbf{T}\left(\phi,\chi\right)^{\text{T}}
  + \frac{1}{2}\left(\phi,\chi\right)\mathbf{M}^2\left(\phi,\chi\right)^{\text{T}} +
  \begin{pmatrix}
    \phi^-_1,\phi^-_2 
  \end{pmatrix}
  \mathbf{M}_{\phi^\pm}^2
  \begin{pmatrix}
    \phi^+_1
    \\
    \phi^+_2
  \end{pmatrix} + \cdots.
\end{align}
\noindent
Here   $\mathbf{T}   =   \left(T_{\phi_1},   T_{\phi_2},   T_{\phi_s},
T_{\chi_1},   T_{\chi_2},  T_{\chi_s}\right)$   denotes  the   tadpole
coefficients  of  the neutral  Higgs  fields,  and $\mathbf{M}^2$  and
$\mathbf{M}_{\phi^\pm}^2$ denote the mass  matrices of the neutral and
charged Higgs bosons, respectively.

Since $\mathbf{M}^2$ and $\mathbf{M}^2_{\phi^\pm}$ are symmetric and
hermitian matrices, respectively, we diagonalize them by an orthogonal
$(6\times 6)$ matrix $\mathbf{U}_{n}$ and a unitary $(2\times 2)$
matrix $\mathbf{U}_{c}$, respectively,
\begin{subequations}
  \begin{align}
    \mathbf{D}_{hG}
    =
    \text{diag}{\left(m_{h_1}^2,\,m_{h_2}^2,\,m_{h_3}^2,\,m_{h_4}^2,\,m_{h_5}^2,\,0\right)}
    &=
    \mathbf{U}_{n}\,\mathbf{M}^2\,\mathbf{U}_{n}^{\textrm{T}}\,,\\
    \mathbf{D}_{h^\pm G^\pm}
    =
    \text{diag}{\left(M_{H^\pm}^2,\,0\right)}
    &=
    \mathbf{U}_{c}\,\mathbf{M}^2_{\phi^\pm}\,\mathbf{U}_{c}^\dagger\,.
  \end{align}
\end{subequations}
\noindent
These  transformations  define  the  five  neutral  Higgs  boson  mass
eigenstates, $h_i$, ($i=1,\ldots,5$), and the (would-be) Goldstone boson
$G$, as  well as  the charged Higgs  and (would-be)  Goldstone states,
$H^\pm$ and $G^\pm$, at the tree level,
\begin{subequations}
  \begin{align}
    \left(h, G\right)^{\textrm{T}} \equiv
    \left(h_1, h_2, h_3, h_4, h_5, G\right)^{\textrm{T}} &=
    \mathbf{U}_{n} \left(\phi,\chi\right)^{\textrm{T}},
    \label{eq:MEStates}\\
    \left(H^\pm, G^\pm\right)^{\textrm{T}}
    &=
    \mathbf{U}_{c} \left(\phi_1^\pm, \phi_2^\pm\right)^{\textrm{T}}.
    \label{eq:MECtates}
  \end{align}
\end{subequations}
\noindent
It is convenient to decompose $\mathbf{U}_{n}$ into two matrices
$\mathbf{U}_{n}^G$ and $\mathbf{U}_{n}^5$, where
$\mathbf{U}_{n}^G$ singularizes out the neutral Goldstone boson,
\begin{align}
  \label{Goldmode}
  \mathbf{U}_{n} \left(\phi,\chi\right)^{\textrm{T}} =
  \mathbf{U}_{n}^5 \mathbf{U}_{n}^G \left(\phi,\chi\right)^{\textrm{T}} =
  \mathbf{U}_{n}^5 \left(\phi_1, \phi_2,  \phi_s, A, \chi_s, G\right)^{\textrm{T}} =
  \left(h, G\right)^{\textrm{T}}.
\end{align}
\noindent
In  the  \cp-violating NMSSM  the  five  fields~$h_i$ are  in  general
superpositions  of  the  \cp-even  and -odd  components  $\phi_i$  and
$\chi_j$. In the special case of
\begin{align}
  \label{eq:CPcond}
  \sin{\left(\xi - 2\,\xi_s + \phi_\lambda - \phi_\kappa\right)} &= 0
\end{align}
\noindent
\cp-conservation is  restored in the  Higgs sector at the  tree level,
and the neutral mass matrix $\mathbf{M}^2$ becomes block-diagonal with
two $(3\times 3)$ sub-matrices for the \cp-even and -odd entries.

The  five linearly  independent  tadpole coefficients  are related  to
soft-breaking terms and combinations of phases as
\begin{subequations}
  \label{eq:TadpoleExpressions}
  \begin{align}
    \begin{split}
    m_1^2 &=
    -\lvert \mueff\rvert^2 - \frac{1}{2}\,M_Z^2\,\cos{\left(2\beta\right)} - \left(\lvert\lambda\rvert\,v\,\sin{\beta}\right)^2 - \frac{T_{\phi_1}}{\sqrt{2}\,v\,\cos{\beta}}\\
    &\quad + \lvert\mueff\rvert\,\tan{\beta}\left(\lvert A_\lambda\rvert\,\cos{\zeta_2} + \frac{\lvert\kappa\rvert\,\lvert\mueff\rvert}{\lvert\lambda\rvert}\cos{\zeta_1}\right)\,,
    \end{split}
    \\
    \begin{split}
    m_2^2 &=
    -\lvert \mueff\rvert^2 + \frac{1}{2}\,M_Z^2\,\cos{\left(2\beta\right)} - \left(\lvert\lambda\rvert\,v\,\cos{\beta}\right)^2 - \frac{T_{\phi_2}}{\sqrt{2}\,v\,\sin{\beta}}\\
    &\quad + \frac{\lvert\mueff\rvert}{\tan{\beta}}\left(\lvert A_\lambda\rvert\,\cos{\zeta_2} + \frac{\lvert\kappa\rvert\,\lvert\mueff\rvert}{\lvert\lambda\rvert}\cos{\zeta_1}\right)\,,
    \end{split}
    \\
    \begin{split}
      m_s^2 &=
      -\lvert\lambda\rvert^2\,v^2 - \frac{T_{\phi_s}\,\lvert\lambda\rvert}{\sqrt{2}\,\lvert\mueff\rvert} - \frac{\lvert\kappa\rvert\,\lvert\mueff\rvert}{\lvert\lambda\rvert}\left(\lvert A_\kappa\rvert\,\cos{\zeta_3} + 2 \frac{\lvert\kappa\rvert\,\lvert\mueff\rvert}{\lvert\lambda\rvert}\right)\\
      &\quad + \frac{1}{2}\frac{\lvert\lambda\rvert^2\,v^2}{\lvert\mueff\rvert}\sin{\left(2\beta\right)}\left(\lvert A_\lambda\rvert\,\cos{\zeta_2} + 2\frac{\lvert\kappa\rvert\,\lvert\mueff\rvert}{\lvert\lambda\rvert}\cos{\zeta_1}\right)
    \end{split}
    \\
    \sin{\zeta_2} &=
    \frac{1}{\lvert A_\lambda\rvert}\left(
    -\frac{\lvert\kappa\rvert\,\lvert\mueff\rvert}{\lvert\lambda\rvert}\sin{\zeta_1}
    - \frac{T_{\chi_1}}{\sqrt{2}\,\lvert\mueff\rvert\,v\,\sin{\beta}}\right)\,,
    \\
    \sin{\zeta_3} &=
    \frac{\lvert\lambda\rvert^2}{\lvert\kappa\rvert\,\lvert\mueff\rvert^2\,\lvert A_\kappa\rvert}\left(
    \frac{\lvert\lambda\rvert\,v^2}{2}\sin{\left(2\beta\right)}\left(\lvert A_\lambda\rvert\,\sin{\zeta_2} - 2\frac{\lvert\kappa\rvert\,\lvert\mueff\rvert}{\lvert\lambda\rvert}\sin{\zeta_1}\right) + \frac{T_{\chi_s}}{\sqrt{2}}
    \right)
    \,,
  \end{align}
\end{subequations}
\noindent
where the masses  of the $W$ and  $Z$ bosons are denoted  by $M_W$ and
$M_Z$,  respectively, and  the  phases  combine to  $\zeta_1  = \xi  -
2\,\xi_s  + \phi_\lambda  - \phi_\kappa$,  $\zeta_2  = \xi  + \xi_s  +
\phi_{A_\lambda}   +  \phi_\lambda$   and   $\zeta_3   =  3\,\xi_s   +
\phi_{A_\kappa}     +     \phi_\kappa$.     The     expressions     of
Eq.~\eqref{eq:TadpoleExpressions}   make   plain  that   the   tadpole
coefficients  can substitute  the  five  parameters $m_1^2$,  $m_2^2$,
$m_S^2$, $\phi_{A_\lambda}$ and $\phi_{A_\kappa}$,  so that the latter
will not  be regarded as  free parameters in the  following.  Finally,
the   tadpole   coefficients   in   the   (tree-level)   mass   basis,
$\mathbf{T}_{h} =  (T_{h_1},T_{h_2},T_{h_3},T_{h_4},T_{h_5},0)$, where
the zero  denotes the vanishing  tadpole coefficient of  the Goldstone
mode, are obtained by $\mathbf{T}_h = \mathbf{U}_{n} \mathbf{T}$.  The
minimization of~$V_{\text{H}}$ at the  chosen Higgs vevs is guaranteed
through  the  condition  that all  tadpole  coefficients  $\mathbf{T}$
vanish at the tree level.
  
The trilinear parameter $\left|A_\lambda\right|$ can be expressed in
terms of the charged Higgs mass $M_{H^\pm}$ as
\begin{align}
  \lvert A_\lambda\rvert\,\cos{\zeta_2} &=
  -\frac{\lvert\kappa\rvert\,\lvert\mueff\rvert}{\lvert\lambda\rvert}\cos{\zeta_1}
  + \left(M_{H^\pm}^2 - M_W^2 + \lvert\lambda\rvert^2\,v^2\right)
  \frac{\sin{\left(2\beta\right)}}{2\,\lvert\mueff\rvert}\,.
\end{align}

Our   renormalization  scheme   will   also   involve  the   fermionic
superpartners of  the Higgs bosons,  known as the higgsinos.   We thus
introduce  here the  Dirac  spinors $\tilde{H}^{\pm}$  of the  charged
higgsino  fields, as  well  as  the Majorana  spinor  of the  singlino
$\tilde{S}$. In  turn, these higgsino  gauge eigenstates mix  with the
gauginos  to  form  the  mass  states known  as  the  neutralinos  and
charginos---see \EG{} Eq.~(11)  in Ref.~\cite{Drechsel:2016jdg}, where
the  NMSSM  parameters  $\lambda$, $\kappa$,  $M_{1,2}$  and  $\mueff$
should be promoted to complex values.  Yet these mass states will play
no role in the discussion below.

\subsection{\label{sec:RenormalisationHiggsPotential}Renormalization of the Higgs potential}

In  the  past,  radiative  corrections  to the  Higgs  masses  of  the
\cp-conserving NMSSM  have been considered in  the effective potential
approach,  see  \EG{} Refs.~\cite{Ellwanger:1992jp,  Ellwanger:1993hn,
  Elliott:1993ex,   Elliott:1993uc,  Elliott:1993bs,   Pandita:1993hx,
  Pandita:1993tg, Ham:1996mi,  Ellwanger:1999ji}. This topic  has also
been  analyzed  from  the  perspective of  a  diagrammatic  expansion,
including  radiative corrections  from part  or  the full  set of  the
particle  content  of   the  NMSSM:  see  Refs.~\cite{Degrassi:2009yq,
  Belanger:2017rgu, Drechsel:2016jdg}.  Both procedures have also been
employed for  the \cp-violating  case: contributions to  the effective
potential   have   been   discussed   in   Refs.~\cite{Matsuda:1995ta,
  Haba:1995aw,      Asatrian:1996wn,     Haba:1996bg,      Ham:1999zs,
  Branco:2000dq,      Davies:2001uv,      Ham:2001wt,      Ham:2001kf,
  Hugonie:2003yu, Ham:2003jf, Funakubo:2004ka, Ham:2007mt, Ham:2008cg,
  Cheung:2010ba,  Moretti:2013lya,   Munir:2013dya,  Domingo:2015qaa},
while  contributions   using  the  diagrammatic  approach   have  been
presented in Refs.~\cite{Graf:2012hh, Goodsell:2014bna}.
  
In the present work, the radiative corrections to the Higgs sector are
calculated  in  the diagrammatic  approach.   To  this end,  we  first
establish a list of the independent parameters appearing in the linear
and bilinear terms of the Higgs potential in Eq.~\eqref{eq:HiggsPot}:
\begin{align}
  \label{renpar}
  T_{h_{1,\ldots,5}}, \ \MHp^2, \ \MW^2, \ \MZ^2, \ e,
  \ \tb, \ \lvert\mueff\rvert, \ \lvert\lambda\rvert,
  \ \phi_{\lambda}, \ \lvert\kappa\rvert, \ \phi_{\kappa}, \ \lvert A_\kappa\rvert, \ \xi, \ \xi_s,
\end{align}
\noindent
where  the  electromagnetic  coupling  $e$  is  related  to  the  fine
structure constant $\alpha$ by $e=\sqrt{4\,\pi\,\alpha}$.  Compared to
Ref.~\cite{Drechsel:2016jdg}, we  use~$e$ as an  independent parameter
instead of the vev~$v$.  This  choice allows to renormalize~$e$ to its
value derived from  the Fermi constant~$G_F$ and does  not require the
reparametrization procedure  employed in Ref.~\cite{Drechsel:2016jdg}.
The  difference between  the  renormalization employed  here, and  the
renormalization   and   reparametrization   procedure   described   in
Ref.~\cite{Drechsel:2016jdg}  is  a  sub-leading  effect  of  two-loop
order,  however.   Other  proposals  in  the   literature  consist  in
fixing~$e$            from~$\alpha\left(M_Z\right)$~\cite{Graf:2012hh,
  Goodsell:2014bna}.

To  all real  and  complex independent  parameters,  $g_r$ and  $g_c$,
respectively,  that  are given  in  Eq.~\eqref{renpar},  we apply  the
renormalization transformations
\begin{align}
  g_r &\rightarrow g_r\left(1 + \delta Z_r\right) = g_r + \delta{g_r}\ ,
  &
  g_c &\rightarrow g_c\left(1 + \delta Z_c\right) =g_c + \delta{g_c} = g_c + \delta{\left|g_c\right|}\,\E^{\I\,\phi_c} + \I\,g_c\,\delta{\phi_c}\ .
\end{align}
\noindent
The  renormalization  transformations  for  the  Higgs,  singlino  and
charged higgsino fields read
\begin{subequations}
  \begin{align}
    \label{eq:dZHiggsN}
    \mathcal{H}_{1, 2} &\rightarrow
    \left(1 + \tfrac{1}{2}\,\delta Z_{\mathcal{H}_{1,2}}\right)\mathcal{H}_{1,2},&
    \mathcal{S} &\rightarrow
    \left(1 + \tfrac{1}{2}\,\delta Z_{\mathcal{S}}\right)\mathcal{S},\\
    \label{eq:dZHiggsC}
    \tilde{H}^\pm &\rightarrow
    \left(1 + \tfrac{1}{2}\,\delta{Z_{\tilde{H}^\pm}^{\textrm{L}}}\,\PrL +  \tfrac{1}{2}\,\delta{Z_{\tilde{H}^\pm}^{\textrm{R}}}\,\PrR\right)\tilde{H}^\pm,&
    \tilde{S} &\rightarrow
    \left(1 + \tfrac{1}{2}\,\delta{Z_{\tilde{S}}^{\textrm{L}}}\,\PrL +\tfrac{1}{2}\,\delta{Z_{\tilde{S}}^{\textrm{R}}}\,\PrR\right) \tilde{S},
  \end{align}
\end{subequations}
\noindent
with $\PrL$ and $\PrR$ denoting the left- and right-handed projectors,
respectively.    Since  the   singlino  is   a  Majorana   field,  the
corresponding                wave-function                counterterms
$\delta{Z_{\tilde{S}}^{\textrm{L}}}$                               and
$\delta{Z_{\tilde{S}}^{\textrm{R}}}$  are  complex conjugates  of  one
another.

\subsection{Renormalization conditions at the one-loop order}

For the  parameters $\MHp$,  $M_W$, $M_Z$ and  $\tb$, which  enter the
one-loop  calculation of  the Higgs  masses in  the MSSM  as well,  we
follow     the     renormalization    prescription     outlined     in
Ref.~\cite{Frank:2006yh}:  the  on-shell   renormalization  scheme  is
employed for the gauge boson masses,  $M_Z$ and $M_W$, and the charged
Higgs  mass   $M_{H^\pm}$,  while   the  parameter   $\tan{\beta}$  is
renormalized \DRbar.

We  apply the  minimization conditions  in  order to  fix the  tadpole
counterterms:
\begin{center}
\vspace{-5ex}
\begin{align}
  T^{(1)}_{h_i} + \delta{T_{h_i}} &= 0,
\end{align}
\vspace{-5ex}
\end{center}
\noindent
where the $T^{(1)}_{h_i}$ correspond  to the one-loop contributions to
the tadpole parameters.

The counterterm of the electromagnetic coupling $e$ is fixed by
\begin{subequations}
  \begin{align}
    \delta{Z_e} &= \delta{Z_e^{\text{Th}}} - \frac{1}{2}\,\Delta{r}^{\textrm{NMSSM}},\\[-1ex]
    \label{eq:ZeThomson}
    \delta Z_e^{\text{Th}} &= \frac{1}{2}\,\Pi^{\gamma\gamma}(0) + \frac{\sw}{\cw}
    \frac{\Sigma_T^{\gamma Z}(0)}{\MZ^2} \ .
  \end{align}
\end{subequations}
\noindent
Here  $\delta  Z_e^{\text{Th}}$  is  the  counterterm  of  the  charge
renormalization  within the  NMSSM according  to the  static (Thomson)
limit.   The quantities  $\Pi^{\gamma\gamma}(0)$ and~$\Sigma_T^{\gamma
  Z}(0)$ are respectively the derivative of the transverse part of the
photon  self-energy  and  the   transverse  part  of  the  photon--$Z$
self-energy  at  zero momentum  transfer.   For  the quantity  $\Delta
r^{\textrm{NMSSM}}$,   relating  the   elementary  charge  to   the  Fermi
constant~$G_F$  measured  in  muon  decays,   we  use  the  result  of
Ref.~\cite{Stal:2015zca}  (see also  Ref.~\cite{Domingo:2011uf}).  The
numerical  value   for  the  electromagnetic  coupling   $e$  in  this
parametrization is obtained  from the Fermi constant in  the usual way
as $e = 2\,\MW\,\sw\,\sqrt{\sqrt{2}\,G_F}$.   This choice differs from
previous works, where either  the charge renormalization condition was
determined  in terms  of $\alpha(M_Z)$~\cite{Graf:2012hh},  or instead
$v$  was  renormalized  \DRbar{}   and  the  result  was  subsequently
reparametrized  to  use  the  value  of~$e$  derived  from  the  Fermi
constant~\cite{Drechsel:2016jdg}.

The  remaining independent  parameters and  the field  renormalization
constants are renormalized \DRbar.   We present a detailed description
of the \DRbar\  renormalization conditions that we  apply.  The actual
cancellation  of UV-divergences,  that we  recover at  the diagrammatic
level,  represents  a  non-trivial  check  for  the  validity  of  the
\FA\ model-file employed for our calculation.

\medskip

The \DRbar\ field  renormalization constants for the  Higgs fields are
obtained as
\begin{align}
  \delta Z_{\mathcal{H}_1} &=
  -\Real{\frac{d\Sigma_{\phi_1\phi_1}^{(1)}}{dp^2}}_{\textrm{div}}
  ,&
  \delta Z_{\mathcal{H}_2} &=
  -\Real{\frac{d\Sigma_{\phi_2\phi_2}^{(1)}}{dp^2}}_{\textrm{div}}
  ,&
  \delta Z_{\mathcal{S}} &=
    -\Real{\frac{d\Sigma_{\phi_s\phi_s}^{(1)}}{dp^2}}_{\textrm{div}},
\end{align}
\noindent
where $\Sigma_{ii}^{(1)}$ denotes the self-energy  of field $i$ at the
one-loop order, and the subscript 'div' denotes the UV-divergent piece
(along with  the universal  finite pieces that  are associated  in the
\DRbar{} scheme) of the quantity that  it follows. The result does not
depend on the momentum~$p^2$.

The field renormalization  constants for the charged  higgsino and the
singlino fields are defined by  the following conditions (the momentum
$p^2$ again does not matter)
\begin{align}
  \delta Z_{\tilde{H}^\pm}^{\text{L}} &= -\left.\Sigma_{\tilde{H}^\pm\tilde{H}^\pm}^{\text{vec\,L}\,(1)}\right|_{\textrm{div}}\,,&
  \delta Z_{\tilde{H}^\pm}^{\text{R}} &= -\left.\Sigma_{\tilde{H}^\pm\tilde{H}^\pm}^{\text{vec\,R}\,(1)}\right|_{\textrm{div}}\,,&
    \delta Z_{\tilde{S}}^{\text{L}} &= -\left.\Sigma_{\tilde{S}\tilde{S}}^{\text{vec\,L}\,(1)}\right|_{\textrm{div}}\,,
\end{align}
\noindent
where the self-energies of the  fermion fields are decomposed into the
left and right vector and the scalar contributions,
\begin{align}
  \Sigma_{ff}^{(1)}{\left(p^2\right)}
  &=
  \Sigma_{ff}^{\text{scal}\,(1)}{\left(p^2\right)} + 
  p_\mu \gamma^\mu 
  \left[\PrL\,\Sigma_{ff}^{\text{vec\,L}\,(1)}{\left(p^2\right)}
    + \PrR\,\Sigma_{ff}^{\text{vec\,R}\,(1)}{\left(p^2\right)}\right].
\end{align}

The    renormalization     constants    $\delta{\lvert\lambda\rvert}$,
$\delta{\lvert\kappa\rvert}$,                $\delta{\phi_{\lambda}}$,
$\delta{\phi_{\kappa}}$, $\delta\xi$, $\delta\xi_s$ and $\delta{\lvert
  A_\kappa\rvert}$  are   fixed  by  \DRbar\  conditions   imposed  on
trilinear   vertices   involving   scalar,   \cp-even   Higgs   fields
$\phi_{1,2,s}$,  the singlino  $\tilde{S}$, and  the charged  higgsino
fields $\tilde{H}^\pm$,  in the  interaction basis  in analogy  to the
procedure  outlined  in~\cite{Drechsel:2016ukp}.  The  renormalization
condition   imposed   on   the   renormalized   three-point   function
$\hat{\Gamma}_{ijk}$ for three arbitrary fields $i$, $j$ and $k$ reads
\begin{align}\label{eq:DRbarGamma}
  \hat{\Gamma}_{ijk} &= \Gamma^{(0)}_{ijk} + \Gamma^{(1)}_{ijk} + \delta{\Gamma}_{ijk} 
  \overset{!}{=} \text{finite}\, ,&\Leftarrow&&
  \delta{\Gamma}_{ijk} &= -\left.\Gamma^{(1)}_{ijk}\right|_{\text{div}}\,,
\end{align}
\noindent
where $\Gamma_{ijk}^{(0)}$ and  $\Gamma_{ijk}^{(1)}$ denote the vertex
function  at  the tree-level  and  one-loop  order, respectively,  and
$\delta{\Gamma_{ijk}}$  denotes  the   counterterm.   The  counterterm
$\delta{\Gamma_{ijk}}$  is thus  fixed by  the divergent  part of  the
vertex  function.  The  renormalization constants  of the  independent
parameters   are   subsequently   fixed   by   linear   relations   to
$\delta{\Gamma_{ijk}}$.
\clearpage
\begin{itemize}
\item For  $\delta{\left|\lambda\right|}$ and $\delta{\phi_{\lambda}}$
  we    impose    the    \DRbar\    renormalization    condition    of
  Eq.~\eqref{eq:DRbarGamma}          on          the          vertices
  ${\Gamma_{\tilde{S}\tilde{H}^-\phi_1^+}^{(0)}   =    \lambda}$   and
  ${\Gamma_{\tilde{S}\tilde{H}^+\phi_1^-}^{(0)}    =   \lambda^\ast}$,
  which yields
  \begin{subequations}
    \begin{align}
      \frac{\delta\left|\lambda\right|}{\left|\lambda\right|} &=
      -\frac{1}{2}\left\{
      \left.\frac{\Gamma_{\tilde{S}\tilde{H}^-\phi_1^+}^{(1)}
      }{
        \Gamma_{\tilde{S}\tilde{H}^-\phi_1^+}^{(0)}}\right|_{\text{div}}
      + \frac{1}{2}\left(
      \delta Z_{\tilde{S}} + \delta Z_{\tilde{H}^\pm}^{\text{R}} + \delta Z_{\mathcal{H}_1}
      \right)\right\} + \text{c.\,c.}\,,
      \\
      \delta\phi_\lambda &=
      -\frac{1}{2\I}\left\{
      \left.\frac{\Gamma_{\tilde{S}\tilde{H}^-\phi_1^+}^{(1)}
      }{
        \Gamma_{\tilde{S}\tilde{H}^-\phi_1^+}^{(0)}}\right|_{\text{div}}
      + \frac{1}{2}\left(
      \delta Z_{\tilde{S}} + \delta Z_{\tilde{H}^\pm}^{\text{R}} + \delta Z_{\mathcal{H}_1}
      \right)\right\} + \text{c.\,c.}
    \end{align}
  \end{subequations}
\item  For  $\delta\xi$ we  impose  the  renormalization condition  of
  Eq.~\eqref{eq:DRbarGamma}          on          the          vertices
  ${\Gamma_{\tilde{S}\tilde{H}^+\phi_2^-}^{(0)}                      =
    \lambda\,\E^{\I\,\xi}}$                                        and
  ${\Gamma_{\tilde{S}\tilde{H}^-\phi_2^+}^{(0)}                      =
    \lambda^\ast\,\E^{-\I\,\xi}}$. The counterterm reads
  \begin{align}
    \delta\xi &=
    \left\{-\frac{1}{2\I}\left[
    \left.\frac{\Gamma_{\tilde{S}\tilde{H}^+\phi_2^-}^{(1)}
    }{
      \Gamma_{\tilde{S}\tilde{H}^+\phi_2^-}^{(0)}}\right|_{\text{div}}
    + \frac{1}{2}\left(
    \delta Z_{\tilde{S}} + \delta Z_{\tilde{H}^\pm}^{\text{L}} + \delta Z_{\mathcal{H}_2}
    \right)\right] + \text{c.\,c.}\right\} - \delta\phi_\lambda\,.
  \end{align}
\item We fix the renormalization  constant $\delta\xi_s$ for the phase
  $\xi_s$  by  applying   Eq.~\eqref{eq:DRbarGamma}  on  the  vertices
  ${\Gamma_{\tilde{H}^-,           \tilde{H}^+\phi_s}^{(0)}          =
  \lambda\,\E^{\I\,\xi_s}/\sqrt{2}}$     and    ${\Gamma_{\tilde{H}^+,
    \tilde{H}^-\phi_s}^{(0)}         =         \lambda^\ast\,\E^{-\I\,
    \xi_s}/\sqrt{2}}$, which yields
  \begin{align}
    \delta\xi_s  &=
    \left\{-\frac{1}{2\I}\left[
    \left.\frac{\Gamma_{\tilde{H}^-, \tilde{H}^+\phi_s}^{(1)}
    }{
      \Gamma_{\tilde{H}^-, \tilde{H}^+\phi_s}^{(0)}}\right|_{\text{div}}
    + \frac{1}{2}\left(
    \delta Z_{\tilde{H}^\pm}^{\text{L}} + \delta Z_{\tilde{H}^\pm}^{\text{R}} + \delta Z_{\phi_s}
    \right)\right] + \text{c.\,c.} \right\} - \delta\phi_\lambda\,.
  \end{align}
\item The  absolute value  and phase of  $\kappa$ are  renormalized by
  $\delta{\left|\kappa\right|}$  and  $\delta{\phi_{\kappa}}$. We  fix
  both renormalization constants by applying Eq.~\eqref{eq:DRbarGamma}
  on    the    vertex   ${\Gamma_{\tilde{S}\tilde{S}\phi_s}^{(0)}    =
    \sqrt{2}\,\kappa\,\E^{\I\,\xi_s}}$, which yields
  \begin{subequations}
    \begin{align}
      \frac{\delta\left|\kappa\right|}{\left|\kappa\right|} &=
      -\frac{1}{2}\left\{
      \left.\frac{\Gamma_{\tilde{S}\tilde{S}\phi_s}^{(1)}
      }{
        \Gamma_{\tilde{S}\tilde{S}\phi_s}^{(0)}}\right|_{\text{div}}
      + \frac{1}{2}\left(
      2\delta Z_{\tilde{S}} + \delta Z_{\phi_s}
      \right)\right\} + \text{c.\,c.}\,,
      \\
      \delta\phi_\kappa &=
      \left\{-\frac{1}{2\I}\left[
      \left.\frac{\Gamma_{\tilde{S}\tilde{S}\phi_s}^{(1)}
      }{
        \Gamma_{\tilde{S}\tilde{S}\phi_s}^{(0)}}\right|_{\text{div}}
      + \frac{1}{2}\left(
      2\delta Z_{\tilde{S}} + \delta Z_{\phi_s}
      \right)\right] + \text{c.\,c.} \right\} - \delta\xi_s\,.
    \end{align}
  \end{subequations}
\item The parameter $\lvert\mueff\rvert$  could be renormalized in the
  on-shell     scheme    for     one    of     the    charginos     or
  neutralinos~\cite{Fritzsche:2011nr,               Heinemeyer:2011gk,
    Bharucha:2012re}.  However, such schemes cannot be stabilized over
  the whole parameter space (due to mass-crossings). We thus prefer to
  apply the \DRbar\ condition
  \begin{align}
    \delta\lvert\mueff\rvert &= \lvert\mueff\rvert\left(\frac{\delta\lvert\lambda\rvert}{\lvert\lambda\rvert}+\frac{1}{2}\delta Z_S\right)\,.
  \end{align}
\item In order to fix $\delta{\left|A_{\kappa}\right|}$, we impose the
  renormalization condition of Eq.~\eqref{eq:DRbarGamma} on the vertex
  $\Gamma_{\phi_s\phi_s\phi_s}^{(0)}  =  -\sqrt{2}\,\lvert\kappa\rvert
  \left(
  \frac{6\,\lvert\kappa\rvert\,\lvert\mueff\rvert}{\lvert\lambda\rvert}
  + \lvert A_\kappa\rvert\cos{\zeta_3}\right)$, ($\zeta_3$ was defined
  after Eq.~\eqref{eq:TadpoleExpressions}).  It reads
  {\allowdisplaybreaks
  \begin{subequations}
  \begin{align}
    \begin{split}
    \delta\lvert A_\kappa\rvert &= -\lvert A_\kappa\rvert \left(
      \frac{\delta{\lvert\kappa\rvert}}{\lvert\kappa\rvert} +
      \frac{\delta{\cos{\zeta_3}}}{\cos{\zeta_3}}\right) -
    \frac{6\,\lvert\kappa\rvert\,\lvert\mueff\rvert}{\lvert\lambda\rvert\,\cos{\zeta_3}}\left(\frac{\delta\lvert\mueff\rvert}{\lvert\mueff\rvert}
      - \frac{\delta\lvert\lambda\rvert}{\lvert\lambda\rvert} +
      \frac{2\,\delta\lvert\kappa\rvert}{\lvert\kappa\rvert}\right)\\
      &\quad-
    \frac{\delta\Gamma_{\phi_s\phi_s\phi_s}}{\sqrt{2}\,\lvert\kappa\rvert\cos{\zeta_3}}\,,
    \end{split}\\
    \begin{split}
      \delta\zeta_3 &=
      \left(\frac{\delta\lvert\kappa\rvert}{\lvert\kappa\rvert} +
        \frac{2\,\delta\lvert\lambda\rvert}{\lvert\lambda\rvert} -
        \frac{\delta \lvert\mu_{\text{eff}}\rvert}{\mu_{\text{eff}}} +
        \frac{\delta\sin{\left(2\beta\right)}}{\sin{\left(2\beta\right)}}
        + \frac{\delta\sin{\zeta_1}}{\sin{\zeta_1}} +
        \left.\frac{2\,\delta
            v}{v}\right|_{\text{div}}\right)\cos{\zeta_3}\sin{\zeta_3}
      \\
      &\quad+
      \left[\frac{\delta\Gamma_{\phi_s\phi_s\phi_s}}{\sqrt{2}\,\lvert\kappa\rvert}
        +
        \frac{6\,\lvert\kappa\rvert\,\lvert\mueff\rvert}{\lvert\lambda\rvert}\left(\frac{\delta\lvert\mueff\rvert}{\lvert\mueff\rvert}
          - \frac{\delta\lvert\lambda\rvert}{\lvert\lambda\rvert} +
          \frac{2\,\delta\lvert\kappa\rvert}{\lvert\kappa\rvert}\right)\right]\frac{\sin{\zeta_3}}{\lvert
        A_\kappa\rvert}
      \\
      &\quad + \left(\left.\delta T_{\chi_s}\right|_{\text{div}} -
        \frac{\lvert\lambda\rvert\,v\,\cos{\beta}}{\lvert\mueff\rvert}\left.\delta
          T_{\chi_1}\right|_{\text{div}}\right)\frac{\lvert\lambda\rvert^2\,\cos{\zeta_3}}{
        \sqrt{2}\,\lvert\kappa\rvert\,\lvert A_\kappa\rvert\,\lvert\mueff\rvert^2}\,.
    \end{split}
  \end{align}
  \end{subequations}
}%
  \noindent
  where  we  used  the   one-loop  relation  $\delta{\cos{\zeta_3}}  =
  -\sin{\zeta_3}\,\delta{\zeta_3}$,  and  $\delta  v$  is not an independent counterterm, but a  quantity
  depending on the counterterms to the electroweak parameters
  \begin{align}\label{eq:deltav}
    \delta v &= v \left(\frac{\delta M_W}{M_W} + \frac{\delta s_{\text{w}}}{s_{\text{w}}} - \delta Z_e\right)\,, &
    \delta s_{\text{w}} &= \frac{c_{\text{w}}^2}{s_{\text{w}}}\left(\frac{\delta M_Z}{M_Z} - \frac{\delta M_W}{M_W}\right)\,.
  \end{align}

\end{itemize}

We  performed various  consistency checks  of  our model  file at  the
one-loop order:
\begin{itemize}
\item
all the renormalized Higgs  self-energies are UV-finite, for arbitrary
values of the momentum,
\item
all  the  vertex-diagram  amplitudes  of a  Higgs  state  decaying  to
SM-particles or a pair of charginos/neutralinos are UV-finite,
\item
the   UV-divergences  of   the   counterterms   to  gauge   couplings,
superpotential  parameters  or  soft  terms are  consistent  with  the
corresponding     one-loop      beta     functions      (see     \EG{}
Refs.~\cite{Ellwanger:2009dp,Martin:1993zk}),
\item
in  the  \cp-conserving  limit,   our  parameters  and  couplings  are
identical   to  the   findings   of  a   previously  developed   model
file~\cite{Drechsel:2016jdg},
\item
in the MSSM  limit, we have found  agreement of the values  of all our
couplings with their  counterparts in the complex  MSSM, obtained with
the model file of~\cite{Fritzsche:2013fta}.
\item     we    checked     that    $\phi_{\lambda}+\xi+\xi_s$     and
  $\phi_{\kappa}+3\,\xi_s$ were the only  relevant combinations of the
  phases $\phi_{\lambda}$,  $\phi_{\kappa}$, $\xi$ and $\xi_s$  at the
  level of amplitudes,
\item
finally,    we    checked    explicitly,   that    the    counterterms
$\delta{\phi_\lambda}$,    $\delta{\phi_\kappa}$,   $\delta\xi$    and
$\delta\xi_s$  vanish when  all NMSSM  contributions are  included, as
pointed out in Ref.~\cite{Graf:2012hh}. This can also be placed in the
perspective                           of                           the
$\beta$-functions~\cite{Cheng:1973nv,Machacek:1983fi,Machacek:1984zw,Jones:1984cx,West:1984dg,Martin:1993zk}:
the phases from the superpotential parameters have no scale-dependence
(at least up to two-loop order);  since $\xi$ and $\xi_s$ are spurious
degrees of freedom, we could  expect their counterterms to present the
same    vanishing    behaviors    as    $\delta{\phi_\lambda}$    and
$\delta{\phi_\kappa}$.
\end{itemize}

\subsection{Quark Yukawa couplings}

The Yukawa  couplings of the top  and bottom quarks, $Y_t$  and $Y_b$,
have a sizable impact on radiative corrections to the Higgs masses. We
present our prescriptions in this subsection.

The top Yukawa  coupling $Y_t=\sqrt{2\sqrt{2}\,G_F}\,m_t/\sin\beta$ is
defined by the on-shell top mass $m_t$.

For the bottom quark, we employ the running \DRbar\ bottom-mass of the
SM  (containing one-loop  QCD corrections),  $\overline{m}_b$, at  the
scale  $m_t$~\cite{Williams:2011bu}.   Additionally, we  subtract  the
possibly   large   $\tan\beta$-enhanced  one-loop   contributions   to
$\overline{m}_b$---induced  by  gaugino--squark  and  higgsino--squark
loops---from  the numerical  definition of  $Y_b$ at  the tree  level:
$Y_b=\sqrt{2\sqrt{2}\,G_F}\,\overline{m}_b/[\cos\beta\,\lvert
  1+\Delta_b  \rvert]$,   where  $\Delta_b$  is  discussed   in  \EG{}
Refs.~\cite{Banks:1987iu,        Hall:1993gn,        Hempfling:1993kv,
  Carena:1994bv,    Carena:1999py,   Eberl:1999he,    Williams:2011bu,
  Baglio:2013iia}.

\subsection{\label{sec:HiggsMassesho}Higgs masses at higher orders}

The masses of the Higgs bosons  are obtained from the complex poles of
the  full  propagator  matrix.    After  rotating  out  the  Goldstone
mode\footnote{Besides  Higgs--$G$  mixing,   we  neglect  the  kinetic
  Higgs--$Z$  and Higgs--photon  mixing,  since  they are  sub-leading
  effects of  two- and  three-loop order, respectively.}   the inverse
propagator matrix for the five Higgs  fields $h_i$ is a $(5 \times 5)$
matrix that reads
\begin{align}\label{eq:ellmasslag}
  \mathbf{\hat{\Delta}}_{hh}^{-1}{\left(k^2\right)}
  =
  i
  \left[k^2\mathbf{1} - 
    \mathbf{D}_{hh}
    + \mathbf{\hat{\Sigma}}_{hh}{\left(k^2\right)}
    \right].
\end{align}
\noindent
Here                         $\mathbf{D}_{hh}                        =
\text{diag}\{m^2_{h_1},\,m^2_{h_2},\,m^2_{h_3},\,m^2_{h_4},\,m^2_{h_5}\}$
denotes the diagonalized  mass matrix of the Higgs  fields without the
Goldstone at the tree  level, and $\mathbf{\hat{\Sigma}}_{hh}$ denotes
the matrix of the renormalized  self-energy corrections of the neutral
Higgs fields.

The five  complex poles of the  propagator are given by  the values of
the squared external  momentum $k^2$ for which the  determinant of the
inverse propagator matrix vanishes,
\begin{align}
  \det{\left[
    \mathbf{\hat{\Delta}}^{-1}_{hh}{\left(k^2\right)}
    \right]_{k^2 = \mathcal{M}^2_i}} &\overset{!}{=} 0\,, &
  \mathcal{M}_i^2 &\overset{!}{=} M_{h_i}^2 + \I\,\Gamma_{h_i}\,M_{h_i}\,, &
  i &\in \{1,\ldots,5\}\,,
\end{align}
\noindent
where  we  have explicitly  stated  the  connection between  the  pole
$\mathcal{M}_i^2$,  the  Higgs  mass  $M_{h_i}$ and  the  total  width
$\Gamma_{h_i}$ for each Higgs field $h_i$.

In  order to  account for  the  imaginary parts  of the  poles of  the
propagator matrix,  we perform  an expansion  of the  self-energies in
terms of  the imaginary part of  the momentum, which is  assumed to be
small (also see section 4.3.5 of~\cite{Williams:2008}),
\begin{align}
  \mathbf{\hat{\Sigma}}_{hh}{\left(k^2\right)}
  \approx
  \mathbf{\hat{\Sigma}}_{hh}{\left(\Real{k^2}\right)} +
  \I\,\Imag{k^2}\,\frac{\mathrm{d}}{\mathrm{d}k^2}\mathbf{\hat{\Sigma}}_{hh}{\left(\Real{k^2}\right)}
\end{align}
\noindent
In       this      work,       the      renormalized       self-energy
$\mathbf{\hat{\Sigma}}_{hh}$,
\begin{align}
  \label{eq:SEapprox}
  \mathbf{\hat{\Sigma}}_{hh}{\left(k^2\right)}
  \approx
  \left.
  \mathbf{\hat{\Sigma}}^{(\text{1L})}_{hh}{\left(k^2\right)}
  \right|^{\text{NMSSM}} +
  \left.
  \mathbf{\hat{\Sigma}}^{(\text{2L})}_{hh}{\left(k^2\right)}
  \right|_{k^2\;=\;0}^{\text{MSSM}}.
\end{align}
\noindent
is evaluated  by taking into  account the full contributions  from the
\cp-violating NMSSM  at one-loop order  and, as an  approximation, the
MSSM-like      contributions      at       two-loop      order      of
$\mathcal{O}{\left(\alpha_t\alpha_s\right)}$~\cite{Heinemeyer:2007aq}
and     $\mathcal{O}{\left(\alpha_{t}^2\right)}$~\cite{Hollik:2014wea,
  Hollik:2014bua}  at vanishing  external momentum  as implemented  in
\texttt{FeynHiggs}.\footnote{Additional  MSSM-like  contributions,  at
  two-loop order or beyond---\EG{} resummation of large logarithms for
  heavy sfermions~\cite{Bahl:2016brp}---could be  incorporated as well
  (see~\cite{Drechsel:2016jdg}).    However,  we   will  confine   our
  discussion in this paper to the leading two-loop contributions.}
  
We note  that the two-loop $O(\alpha_b\alpha_s)$  contributions to the
Higgs self-energies are not included  in our calculation. Still, as we
employ the  running bottom  mass in the  definition of  $Y_b$ entering
$\left.\mathbf{\hat{\Sigma}}^{(\text{1L})}_{hh}{\left(k^2\right)}
\right|^{\text{NMSSM}}$, we expect that  the missing two-loop piece is
numerically
subleading~\cite{Dedes:2003km,Heinemeyer:2004xw,Heinemeyer:2010mm}.

\subsection[Wave function normalization factors: the matrix $\mathbf{Z}^{\mbox{\tiny mix}}$]{\boldmath Wave function normalization factors: the matrix $\mathbf{Z}^{\mbox{\tiny mix}}$}

In the Feynman-diagrammatic approach  physical processes with external
Higgs fields are defined in terms of the tree-level mass states $h_i$.
When   higher-order  contributions   are   considered,  however,   the
tree-level  mass states  are  not physical  states. Indeed,  radiative
corrections  induce additional  mass  and kinematic  mixing among  the
fields  $h_i$, and  the poles  of  the tree-level  propagators do  not
coincide with  $\mathcal{M}_i^2$.  A  relation between  the amplitudes
with  an  external tree-level  Higgs  mass  state  and those  with  an
external physical  Higgs state is  necessary (though this  relation is
trivial if the fields are  renormalized on-shell).  For example, for a
Higgs decaying into two fermions $f$ this relation is given by the LSZ
reduction formula,
\begin{align}
  \label{eq:RelationPhysAmplitude}
  {\cal   A}[h_i^{\mbox{\tiny  phys}}\to   f\bar{f}]=Z^{\mbox{\tiny
      mix}}_{ij}\,{\cal A}[h_j\to f\bar{f}].
\end{align}
\noindent
Here the  superscript 'phys'  denotes the  amplitude with  an external
physical field.   The coefficients $Z^{\mbox{\tiny mix}}_{ij}$  can be
expressed  explicitly in  terms  of the  full  propagator matrix  (see
Refs.~\cite{Chankowski:1992er,    Heinemeyer:2001iy,    Fuchs:2016swt,
  Fuchs:2017wkq} and also section~$5.3$ of Ref.~\cite{Fuchs:2015jwa}),
as
\begin{align}
  \label{eq:ZLSZ}
  Z^{\mbox{\tiny mix}}_{ij}
  =
  \left.\left[
    \I\,\frac{\mathrm{d}}{\mathrm{d}k^2}\left(\mathbf{\hat{\Delta}}_{hh}^{-1}(k^2)\right)_{ii}
    \right]^{\frac{1}{2}}
  \frac{
    \left(\mathbf{\hat{\Delta}}_{hh}(k^2)\right)_{ij}
  }{
    \left(\mathbf{\hat{\Delta}}_{hh}(k^2)\right)_{ii}}
  \right|_{k^2\to \mathcal{M}^2_i}.
\end{align}
\noindent
However, the  analytical inversion becomes time-consuming  in the case
of     a     $(5\times5)$    propagator     matrix.      Additionally,
$\mathbf{\hat{\Delta}}_{hh}$ needs  to be  evaluated at (or  close to)
its  singular points  $\mathcal{M}_i^2$, which  can lead  to numerical
instabilities on the right-hand  side of Eq.~\eqref{eq:ZLSZ} (only the
ratio          of          propagator         matrix          elements
$(\mathbf{\hat{\Delta}}_{hh})_{ij}$  is finite).   In  order to  avoid
these issues we employed an equivalent formulation of the coefficients
$Z^{\mbox{\tiny mix}}_{ij}$, which is outlined below.

We consider the following effective Lagrangian for the tree-level mass
states $h_i^{\mbox{\tiny (tree)}}$,
\begin{align}
  {\cal L}_{\mbox{\tiny eff}}
  =
  \frac{1}{2\I}\left[
    \mathbf{\hat{\Delta}}^{-1}_{hh}{\left(k^2\right)}\right]_{ij}h_i^{\mbox{\tiny (tree)}}h_j^{\mbox{\tiny (tree)}}
\end{align}
\noindent
and set $\mathbf{Z}^{\mbox{\tiny  mix}} = (Z^{\mbox{\tiny mix}}_{ij})$
as   the   transition   matrix   to   the   physical   Higgs   fields:
$h_i^{\mbox{\tiny  (loop)}}=Z^{\mbox{\tiny  mix}}_{ij}h_j^{\mbox{\tiny
    (tree)}}$.   The states  $h_i^{\mbox{\tiny  (loop)}}$ are  defined
such that
\begin{align}\label{physcond}
 \mathcal{L}_{\mbox{\tiny eff}} = \frac{1}{2}\left[k^2 - \mathcal{M}_{h_i}^2\right](h_i^{\mbox{\tiny (loop)}})^2 + \mathcal{O}{\left(\left[k^2 - \mathcal{M}_{h_i}^2\right]^2\right)}\,.
\end{align}
\noindent
In  other  terms, the  $h_i^{\mbox{\tiny  (loop)}}$  should appear  as
on-shell fields with standard kinetic  terms close to their mass-pole.
Thus the  coefficients $Z^{\mbox{\tiny mix}}_{ij}$ should  satisfy the
`eigenvalue'  conditions\footnote{Note that  the system  is non-linear
  due  to  the  momentum dependence  of  $\mathbf{\hat{\Sigma}}_{hG}$.
  However,   $\left\{\mathcal{M}_{h_i}^2,\left(\mathbf{Z}^{\mbox{\tiny
      mix}}\right)_{i}\right\}$   is    a   genuine    eigenstate   of
  $\mathbf{D}_{hG}                                                   -
  \mathbf{\hat{\Sigma}}_{hG}{\left(\mathcal{M}_{h_i}^2\right)}$.}
\begin{align}\label{physeigen}
 \left[\mathbf{D}_{hG} - \mathbf{\hat{\Sigma}}_{hG}{\left(\mathcal{M}_{h_i}^2\right)}\right]_{kl}Z^{\mbox{\tiny mix}}_{il} &= \mathcal{M}_{h_i}^2Z^{\mbox{\tiny mix}}_{ik}\,.
\end{align}
\noindent
Once                   the                  roots                   of
$\det{\left[\mathbf{\hat{\Delta}}^{-1}_{hh}{\left(k^2\right)}\right]}$
are known, the $i$-th line  of $\mathbf{Z}^{\mbox{\tiny mix}}$ is thus
determined    as    the     eigenvector    of    $\mathbf{D}_{hG}    -
\mathbf{\hat{\Sigma}}_{hG}$ for the eigenvalue $\mathcal{M}_{h_i}^2$.

Finally,    the    normalization    of     the    $i$-th    line    of
$\mathbf{Z}^{\mbox{\tiny   mix}}$  is   specified  by   the  following
condition on the kinetic term,
\begin{align}\label{physnorm}
  \left[
    \frac{\mathrm{d} \mathbf{\hat{\Delta}}^{-1}_{hh}}{\mathrm{d}k^2} {\left(\mathcal{M}_i^2\right)}
    \right]_{kl} Z^{\mbox{\tiny mix}}_{ik}\,Z^{\mbox{\tiny mix}}_{il}
  =  
  \left[\mathbb{I} + \frac{\mathrm{d}\mathbf{\hat{\Sigma}}_{hG}}{
      \mathrm{d}k^2}{\left(\mathcal{M}_i^2\right)}\right]_{kl} Z^{\mbox{\tiny mix}}_{ik}\,Z^{\mbox{\tiny mix}}_{il}
  = 1,
\end{align}
\noindent
such that  the coefficients  $Z^{\mbox{\tiny mix}}_{ij}$  are uniquely
specified   (up  to   a   sign  without   physical  meaning)   by
Eq.~\eqref{physcond}.   For  the  study   of  effects  from  the
  normalization of  $\mathbf{Z}^{\mbox{\tiny mix}}$, it  is convenient
  to define the (squared) norm $\left|Z_i\right|^2$ of its rows,
\begin{align}
  \label{eq:norm}
  \lvert Z_i\rvert^2 = \sum_{j\;=\;1}^{5}{\left|Z^{\mbox{\tiny     mix}}_{ij}\right|^2}\,,
\end{align}
\noindent
\IE{} $\left|Z_i\right|$  correspond to the norm  of the eigenvectors,
that  are  associated to  the  complex  pole $\mathcal{M}_{h_i}$,  see
Eqs.~\eqref{physcond} and~\eqref{physnorm}.

The determination  of $\mathbf{Z}^{\mbox{\tiny mix}}$ in  terms of the
eigenstates of  $\mathbf{\hat{\Delta}}^{-1}_{hh}{\left(k^2\right)}$ is
numerically   easier   to   handle    than   its   determination   via
Eq.~\eqref{eq:ZLSZ}.    Applying    the   two    defining   conditions
Eqs.~\eqref{physeigen}  and  \eqref{physnorm}  to  the  expression  of
Eq.~\eqref{eq:ZLSZ},  one   can  verify   that  both   definitions  of
$\mathbf{Z}^{\mbox{\tiny mix}}$ are identical.

As we discussed, the components of the matrix $\mathbf{Z}^{\mbox{\tiny
    mix}}$ establish  the connection  between the physical  fields and
the  tree-level   external  legs.   In  the   literature  this  matrix
$\mathbf{Z}^{\mbox{\tiny  mix}}$  is   often  replaced  by  simplified
versions neglecting the momentum dependence of the self-energies. With
the aim of performing numerical  comparisons in the following section,
we introduce two such approximate  definitions of the relation between
loop-corrected and tree-level fields:
\begin{itemize}
 \item The first approach consists in freezing the momentum to $k^2=0$
   in the self-energy of Eq.~\eqref{eq:ellmasslag}. This assumption is
   known as the effective  potential approximation.  In this
   approach        the         inverse        propagator        matrix
   $\mathbf{\Delta}_{hh}^{-1}{(k^2=0)}$,       as       given       by
   Eq.~\eqref{eq:ellmasslag}, is diagonalized by a simple orthogonal matrix
   $\mathbf{U}^0$,     which    approximates
   $\mathbf{Z}^{\mbox{\tiny mix}}$.
 \item Another choice consists in replacing the momentum dependence of
   the      self-energy      in      Eq.~\eqref{eq:ellmasslag}      by
   $\left[\mathbf{\hat{\Sigma}}_{hh}{\left(k^2\right)}\right]_{ij}\rightarrow
   \left[\mathbf{\hat{\Sigma}}_{hh}{\left((m_{h_i}^2+m_{h_j}^2)/2\right)}\right]_{ij}$
   (given in the basis of  the tree-level mass states). This procedure
   aims  at   more  closely  mimicking   the  actual  values   of  the
   self-energies involved  in the mass calculation.   In this approach
   the inverse propagator is also diagonalized by an orthogonal matrix
   $\mathbf{U}^m$.
\end{itemize}
While these procedures capture the mixing effects induced by radiative
corrections, at least partially, it  is nevertheless obvious that they
miss the normalization of the fields outlined in Eq.~\eqref{physnorm}.
This   means   in   particular   that   the   norm   as   defined   in
Eq.~\eqref{eq:norm}  will  always be  identical  to  1 if  the  matrix
$\mathbf{Z}^{\mbox{\tiny   mix}}$    is   approximated    by   either
$\mathbf{U}^0$ or $\mathbf{U}^m$. We will  discuss the impact of these
approximations in the following section.

\section{\label{sec:numerics}Numerical Analysis}

In this section  we present the results of our  Higgs mass calculation
and  compare  them  with  the  output  of  public  tools  for  several
\cp-violating  scenarios.  We  also investigate  the relevance  of the
matrix  $\mathbf{Z}^{\mbox{\tiny mix}}$  for transition  amplitudes in
the example of the one-loop corrected decays of one Higgs field into a
tau/anti-tau pair, $h_i\to\tau^+\tau^-$.

The choice  for the top-quark  mass is $m_t =  173.2$\,GeV. Throughout
this  section all  \DRbar\ parameters  are defined  at $m_t$,  and all
stop-parameters are on-shell parameters.

From  the  point of  view  of  the  Higgs  phenomenology we  test  the
scenarios presented in this section  with the full set of experimental
constraints   and   signals   implemented    in   the   public   tools
\texttt{HiggsBounds-4.3.1}~\cite{Bechtle:2008jh,       Bechtle:2011sb,
  Bechtle:2013gu,   Bechtle:2013wla,   Bechtle:2015pma,  HB-www}   and
\texttt{HiggsSignals-1.3.1}~\cite{Bechtle:2013xfa, HB-www}.

\subsection{Comparison with \texttt{FeynHiggs} in the MSSM-limit}

In  the  limit  of  vanishing  $\lambda$  and  $\kappa$,  the  singlet
superfield decouples  from the  MSSM sector  and one  is left  with an
effective   MSSM---the    $\mueff$   term   persists   as    long   as
$\kappa\sim\lambda$.  We  may then compare  our results for  the Higgs
masses and the matrix $\mathbf{Z}^{\mbox{\tiny mix}}$ in this limit to
those  of  \verb|FeynHiggs-2.12.0|.  For  this  to  be meaningful,  we
adjust  the  settings of  \verb|FeynHiggs|,  so  that they  match  the
higher-order  contributions and  renormalization scheme  of our  NMSSM
calculation.    In   particular,   we   impose   that   the   one-loop
field-renormalization      constants      and     $\tan\beta$      are
\DRbar-renormalized and select full one-loop and leading two-loop MSSM
contributions of  $\mathcal{O}{(\alpha_t\alpha_s +  \alpha_t^2)}$.  We
also  require  that  \verb|FeynHiggs| takes  the  $\tan\beta$-enhanced
contributions  to the  down-type Yukawa  couplings into  account.  The
corresponding      \texttt{FeynHiggs}       input      flags      read
\texttt{FHSetFlags[4,0,0,3,0,2,0,0,1,1]}.

\begin{figure}[t]
  \includegraphics[width=\linewidth]{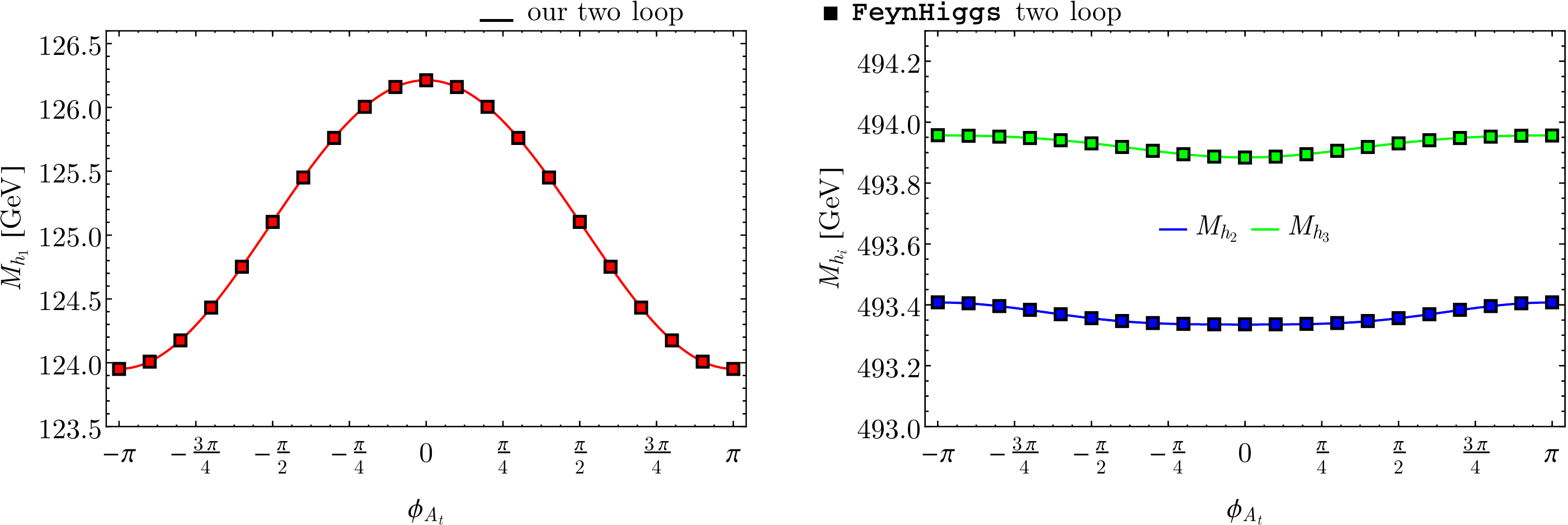}
  \caption{\label{fig:MSSMlim}  Masses  of  the light,  SM-like  state
    $h_1$ (left plot) and the two  heavy states $h_2$ and $h_3$ (right
    plot) as  a function of  $\phi_{A_t}$. The solid lines  denote the
    masses obtained with our calculation, while the squares denote the
    masses obtained with \FH\ with  the options indicated in the text.
    The scenario is representative of  the MSSM-limit of the NMSSM and
    we  employ  the  following  parameters:  $\lambda=\kappa=10^{-5}$,
    $\tan\beta=10$,     $m_{H^{\pm}}=500$\,GeV,     $\mueff=250$\,GeV,
    $A_{\kappa}=-100$\,GeV,                  $m_{\tilde{F}}=1.5$\,TeV,
    $|A_t|=A_b=2.5$\,TeV, $2\,M_1=M_2=M_3/5=0.5$\,TeV.}
\end{figure}

\begin{figure}[b]
  \includegraphics[width=\linewidth]{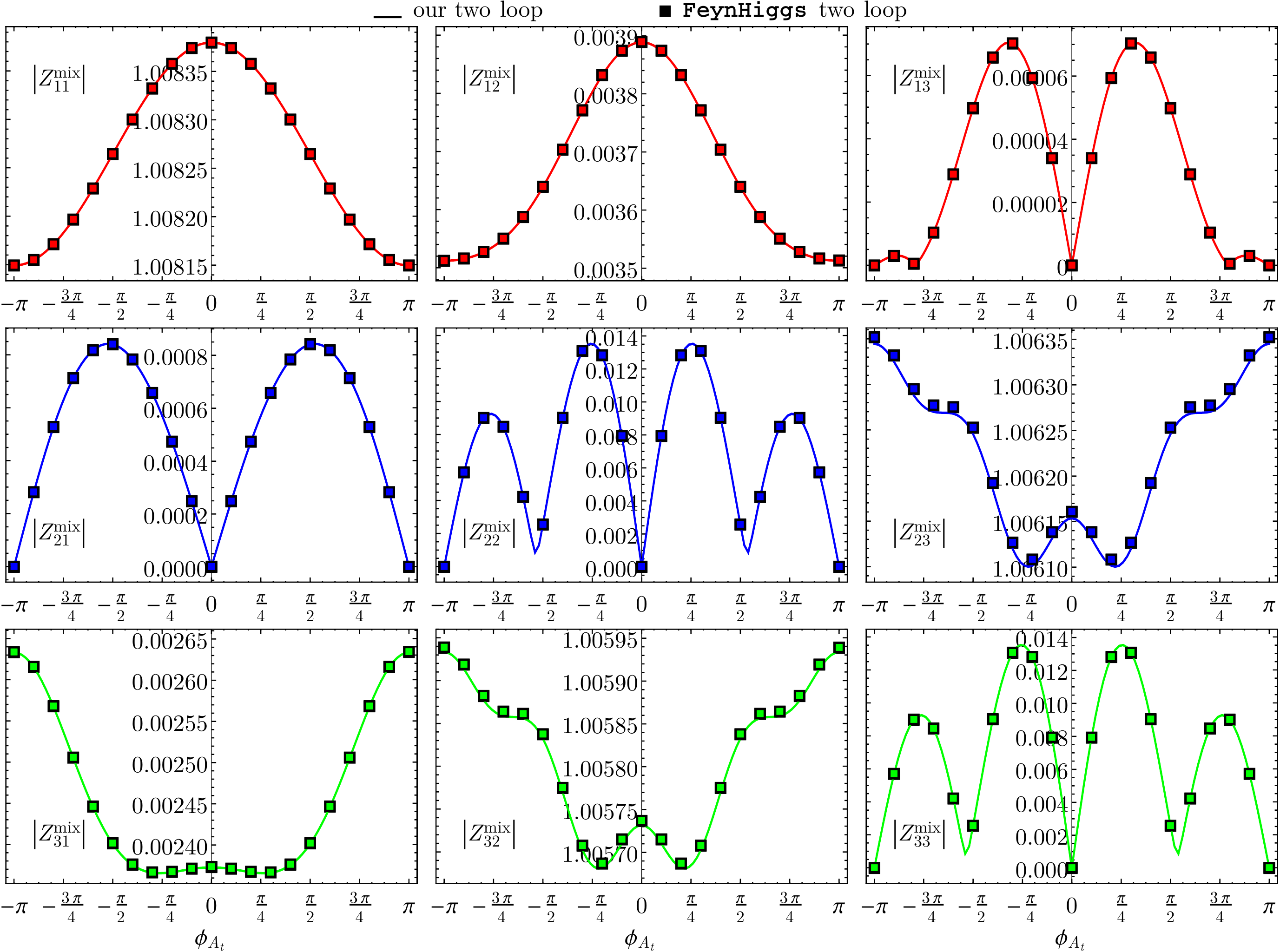}
  \caption{Modules   for    the   nine   elements   of    the   matrix
    $\mathbf{Z}^{\mbox{\tiny  mix}}$,  obtained with  our  calculation
    (solid line) and \FH\ (squares).  The colors follow the convention
    of Fig.~\ref{fig:MSSMlim}:  red for the coefficients  defining the
    wave-function  normalization of  the light,  SM-like state  $h_1$,
    blue and  green for  the coefficients  corresponding to  the heavy
    states $h_2$  and $h_3$, respectively.  The  parameters are chosen
    as in Fig.~\ref{fig:MSSMlim}.}
  \label{fig:MSSMZlim}
\end{figure}

We consider  a region  in the  parameter space of  the NMSSM  with the
following  characteristics: $\lambda=\kappa=10^{-5}$,  $\tan\beta=10$,
$m_{H^{\pm}}=500$\,GeV, $\mueff=250$\,GeV, $A_{\kappa}=-100$\,GeV; the
sfermion soft masses are set to  the universal value of $1.5$\,TeV and
the sfermion  trilinear couplings to  a value of $0.5$\,TeV,  with the
exception of the third generation parameters $|A_t|=A_b=2.5$\,TeV; the
gaugino masses  are chosen as follows:  $2M_1=M_2=M_3/5=0.5$\,TeV.  We
then vary the phase $\phi_{A_t}$.   A variation of $\phi_{A_t}$ (or of
any  MSSM-like  phase)  in  such  a  naive  direction  is  of  limited
phenomenological  interest, since  in this  case limits  from Electric
Dipole Moments  (EDM) are  violated almost  as soon  as \cp,  see e.g.
Ref.~\cite{Arbey:2014msa}.  In  the following  we dismiss  this issue,
however, and allow  $\phi_{A_t}$ to vary over its  full range. Indeed,
we  are  only  interested  in  comparing our  results  with  those  of
\verb|FeynHiggs|. Due to  the largely SM-like properties  of the state
with  a mass  close  to  $125$\,GeV, our  scenario  appears to  retain
characteristics  that  are  compatible   with  the  experimental  data
implemented  in \texttt{HiggsBounds}  and \texttt{HiggsSignals},  over
the whole range of $\phi_{A_t}$.  The results for the Higgs masses are
displayed  in  Fig.~\ref{fig:MSSMlim}.   We  observe  a  near  perfect
agreement   between  our   results   (solid  curves)   and  those   of
\verb|FeynHiggs|  (squares)  with  differences  of  order  MeV.   This
agreement is expected,  since we closely follow the  procedure for the
renormalization  and  processing  of   the  MSSM-like  input  of  \FH.
Moreover,  due  to  the  small  values  for  $\lambda$  and  $\kappa$,
deviations induced  by genuine  NMSSM effects remain  negligible.  The
results for the elements of the matrix $\mathbf{Z}^{\mbox{\tiny mix}}$
are displayed in Fig.~\ref{fig:MSSMZlim}.  Again,  we find a very good
agreement  between  our  results   and  \FH\  with  differences  below
1\permil\ for the modules.

\subsection{Comparison in the \cp-conserving limit}

We now  turn away from  the MSSM limit.   Our mass calculation  can be
confronted to the routines presented in~\cite{Drechsel:2016jdg} in the
\cp-conserving   case.    Both   approaches   employ    an   identical
renormalization  scheme  in this  limit,  with  the exception  of  the
electroweak   vev,   which   receives   a   \DRbar{}   renormalization
in~\cite{Drechsel:2016jdg} while we parametrize $v$ in terms of $M_W$,
$M_Z$     and    $e$     (see    Eq.~\eqref{eq:deltav}).      However,
in~\cite{Drechsel:2016jdg}  the  input  for  $v$  is  obtained  via  a
reparametrization from  our scheme (the scheme  using $\alpha(M_Z)$ as
input  is  also considered),  as  explained  in  section 2.3  of  that
reference.  Therefore,  both mass predictions are  directly comparable
and the  mismatch between them  should be  understood as an  effect of
two-loop  electroweak   order,  due  to  the   approximations  in  the
reparametrization used by~\cite{Drechsel:2016jdg}.

\begin{figure}[t]
  \vspace{2ex}
  \includegraphics[width=\linewidth]{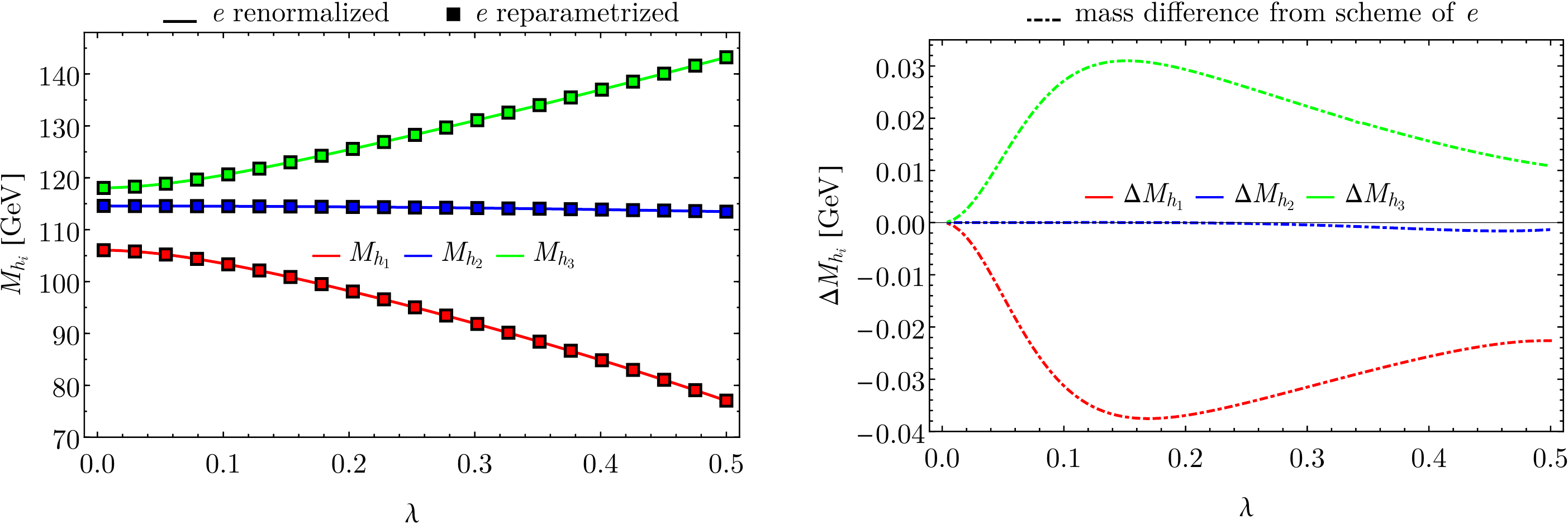}
  \caption{\label{fig:NMSSMreal}Masses  of  the three  lightest  Higgs
    states   $h_1$  (red),   $h_2$  (blue),   $h_3$  (green)   in  the
    \cp-conserving  limit  for  varying  $\lambda=2\,\kappa$  and  the
    following     input:     $\tan\beta=10$,     $M_{H^{\pm}}=1$\,TeV,
    $\mueff=125$\,GeV,                          $A_{\kappa}=-70$\,GeV,
    $m_{\tilde{F}}=1.5$\,TeV, $A_t=2$\,TeV, $A_{f\;\neq\;t}=0.5$\,TeV,
    $2\,M_1=M_2=M_3/5=0.5$\,TeV. On the left-hand side, we present our
    result   (solid   line)   and  that   of   \cite{Drechsel:2016jdg}
    (squares). On  the right-hand  side, we plot  the mass-differences
    between these two codes, due to the scheme applied to the electric
    coupling.}
\end{figure}

We consider the following region in  the parameter space of the NMSSM:
$\kappa=\lambda/2$,        $\tan\beta=10$,       $M_{H^{\pm}}=1$\,TeV,
$\mueff=125$\,GeV, $A_{\kappa}=-70$\,GeV; the soft masses are taken as
in the previous subsection while the trilinear soft sfermion couplings
are all  set to $0.5$\,TeV,  with the exception of  $A_t=1.2$\,TeV. We
scan  over $\lambda$  from $\sim  0$ (the  MSSM-limit) to  $0.5$.  The
masses of the three lightest Higgs states are displayed in the plot on
the left-hand side of Fig.~\ref{fig:NMSSMreal}. The dominantly SM-like
state is  the heaviest of  the three  (green curve).  The  two lighter
states are dominantly  singlet, \cp-even (red curve)  or \cp-odd (blue
curve).   In  the MSSM-limit,  these  three  states are  significantly
lighter   than   $125$~GeV:   this  results   in   an   unsatisfactory
phenomenological  situation in  view  of the  LHC measurements.   With
increasing $\lambda$, the \cp-even  singlet-doublet mixing uplifts the
mass of  the dominantly  SM-like state, leading  to phenomenologically
viable  characteristics---as tested  with \HS---for  $\lambda\sim0.2$.
There, we observe that $h_1$ possesses a sizable doublet component and
a  mass $M_{h_1}\simeq100$\,GeV,  so that  this state  could offer  an
interpretation  of the  local  excess observed  at  LEP in  $e^+e^-\to
Z+(H\to b\bar{b})$ searches~\cite{Barate:2003sz}.

We  then   focus  on   the  comparison   with  the   masses  predicted
by~\cite{Drechsel:2016jdg}.   The  plot  on   the  left-hand  side  of
Fig.~\ref{fig:NMSSMreal} illustrates  a general agreement  between our
calculation (solid curves)  and the results of~\cite{Drechsel:2016jdg}
(squares).   On the  right-hand side  of Fig.~\ref{fig:NMSSMreal},  we
display the mass differences between the two procedures, which are due
to differences of two-loop order induced by the reparametrization used
by~\cite{Drechsel:2016jdg}.   We  observe  vanishing  effects  in  the
MSSM-limit     while      the     mass      differences     eventually
reach~$\mathcal{O}{\left(40\,\text{MeV}\right)}$
for~$\lambda\simeq0.16$.   This can  be  understood  in the  following
fashion: the leading effect originates in the Higgs mass matrix at the
tree   level,   where   an   explicit  dependence   on   $v$   appears
only\footnote{We      remind      the       reader      that      both
  in~\cite{Drechsel:2016jdg}  and  in   our  calculation  $M_W^2$  and
  $M_Z^2$ are chosen as  independent, on-shell parameters.  Therefore,
  the corresponding terms in the Higgs mass-matrix are not affected by
  the differences  in the  renormalization/reparametrization discussed
  here.}  through  terms   of  the  form~$\lambda\,v$  and~$\kappa\,v$
(quadratically for the doublet and  singlet mass entries, and linearly
for  the   doublet--singlet  mixing).    These  terms   are  processed
differently  in  both  approaches:  in~\cite{Drechsel:2016jdg}~$v$  is
regarded  as   an  independent   \DRbar{}  parameter,  while   in  our
calculation~$v$ is a dependent quantity  that is expressed in terms of
the   independent   parameters~$M_W$,   $M_Z$  and~$e$.    While   the
reparametrization   of~\cite{Drechsel:2016jdg}   should  restore   the
agreement between  the two  procedures, neglected effects  of two-loop
electroweak  order  in  this   reparametrization  result  in  a  small
mismatch.   Since  the  terms  that convey  this  mismatch  come  with
prefactors~$\lambda$ or~$\kappa$, the difference  vanishes in the MSSM
limit  ($\lambda,\kappa\to   0$).   Moreover,  in  the   regime  under
consideration, where~$\tan\beta  \gg 1$, it is  possible to understand
why  the  mass  of  the   \cp-odd  singlet  (blue  curve)  is  largely
insensitive  to the  mismatch: terms  $\propto (\lambda\,v)^2$  in the
\cp-odd   singlet   mass   entry  are   suppressed   as~$1/\tan\beta$.
Additionally,      leading     one-loop      radiative     corrections
of~$\mathcal{O}{\left(\alpha_t\right)}$  induce further  dependence on
the processing  of~$v$. However, these corrections  are suppressed for
the points  of Fig.~\ref{fig:NMSSMreal},  as the stops  are relatively
light.

On   the   whole,   the   numerical  mismatch   with   the   procedure
of~\cite{Drechsel:2016jdg}  is very  minor, which  places our  current
code in the direct continuity of this earlier work.

\subsection{Comparison with \NC}

\NC\ is particularly  suitable for a comparison  with our calculation,
since its  mixed \DRbar/on-shell renormalization scheme  is relatively
close to the  one that we use.\footnote{Note that it  is somewhat more
  involved to compare our results quantitatively with RGE-based tools,
  as  the  input  requires  a conversion  to  the  appropriate  scheme
  (usually  \DRbar)  and   a  running  to  the   correct  input  scale
  \cite{Staub:2015aea}.   For  this  reason,   we  shall  confine  our
  discussion   to   comparisons   with  \NC,   which   shares   closer
  characteristics  with our  approach. A  similar comparison  for real
  parameters has been  presented in~\cite{Drechsel:2016htw}.}  Yet, we
note  several differences  between  the  prescriptions implemented  by
\NC\    and    the    procedure    that   we    have    outlined    in
section~\ref{sec:theory}   (defining  our   ``default''  calculation).
First, \NC\ applies  a renormalization scheme for  the electric charge
employing  $\alpha(M_Z)$  as  input---whereas  we  decided  to  define
$\alpha$ via its relation to $G_F$.   Then the input parameters in the
stop  sector are  defined in  the  \DRbar{} scheme  in \NC---while  we
employ on-shell definitions.  Additionally, we resum large $\tan\beta$
effects from  our definition of  the bottom Yukawa, contrarily  to the
Higgs-mass   calculation  of   \NC.   Finally,   \NC\  includes   only
$\mathcal{O}{\left(\alpha_t\alpha_s\right)}$   corrections    at   the
two-loop             order---where             we             consider
$\mathcal{O}{\left(\alpha_t^2\right)}$ effects as  well.  However, the
two-loop $\mathcal{O}{\left(\alpha_t\alpha_s\right)}$ contributions of
\NC\  are  exhaustive in  the  NMSSM  (including corrections  for  the
self-energies with at least one external singlet field)---whereas ours
are obtained in the MSSM approximation.

These observations mean that our  mass-calculation is not directly (at
least,  not  quantitatively) comparable  to  the  predictions of  \NC,
since, of  the items listed above,  the first few produce  a deviation
relative to  the scheme, while the  later ones generate a  mismatch of
higher orders. Consequently, several adjustments  need to be performed
in order  to make a comparison  meaningful and control the  sources of
deviations.   Thus,  \NC\  has  been adjusted  in  view  of  accepting
on-shell  input  in the  stop  sector.\footnote{We  thank K.~Walz  for
  providing a modified  version of \NC\ for  this feature.}  Moreover,
we also establish a ``modified'' version of our routines that attempts
to  mimic   the  choices  of  \NC---\IE{}   employing  $\alpha(M_Z)$,
discarding  large-$\tan\beta$   effects  for  $Y_b$   and  subtracting
$\mathcal{O}{\left(\alpha_t^2\right)}$    corrections---although    we
cannot currently  include $\mathcal{O}{\left(\alpha_t\alpha_s\right)}$
corrections beyond  the MSSM, so  that this effect should  control the
difference of  our modified version  with \NC. Beyond  this comparison
with \NC,  we will  also try  to quantify the  magnitude of  the other
higher-order effects that distinguish our ``default'' result from \NC.

First, we  consider the regime of  the NMSSM with low  $\tan\beta$ and
large  $\lambda$. This  region in  parameter space  is well-known  for
maximizing the specific NMSSM tree-level  contributions to the mass of
the SM-like  Higgs state as  well as for  stimulating singlet--doublet
mixing effects and  other genuine aspects of  the NMSSM phenomenology.
We  employ the  following  parameters: $\lambda=0.7$,  $|\kappa|=0.1$,
$\tan\beta=2$,       $M_{H^{\pm}}=1170$\,GeV,       $\mueff=500$\,GeV,
$A_{\kappa}=-70$\,GeV;   the    soft   masses   are   taken    as   in
Fig.~\ref{fig:MSSMlim} with the exception of  the squarks of the third
generation, for which the soft  masses and trilinear couplings are set
to  $500$\,GeV  and $100$\,GeV,  respectively.   In  the regime  under
consideration genuine NMSSM effects are indeed sufficient to produce a
SM-like  state  in the  observed  mass-range  without requiring  large
top/stop corrections.  We vary  the phase $\phi_{\kappa}$ (we restrict
to a range where the tree-level squared Higgs masses remain positive).
We  note that,  contrarily to  MSSM-like phases,  the phases  from the
singlet  sector  are  allowed  a   wide  range  of  variation  without
conflicting with the measured EDM~\cite{King:2015oxa,Cheung:2011wn}.

\begin{figure}[t]
  \vspace{2ex}
  \begin{center}
  \includegraphics[width=.5\linewidth]{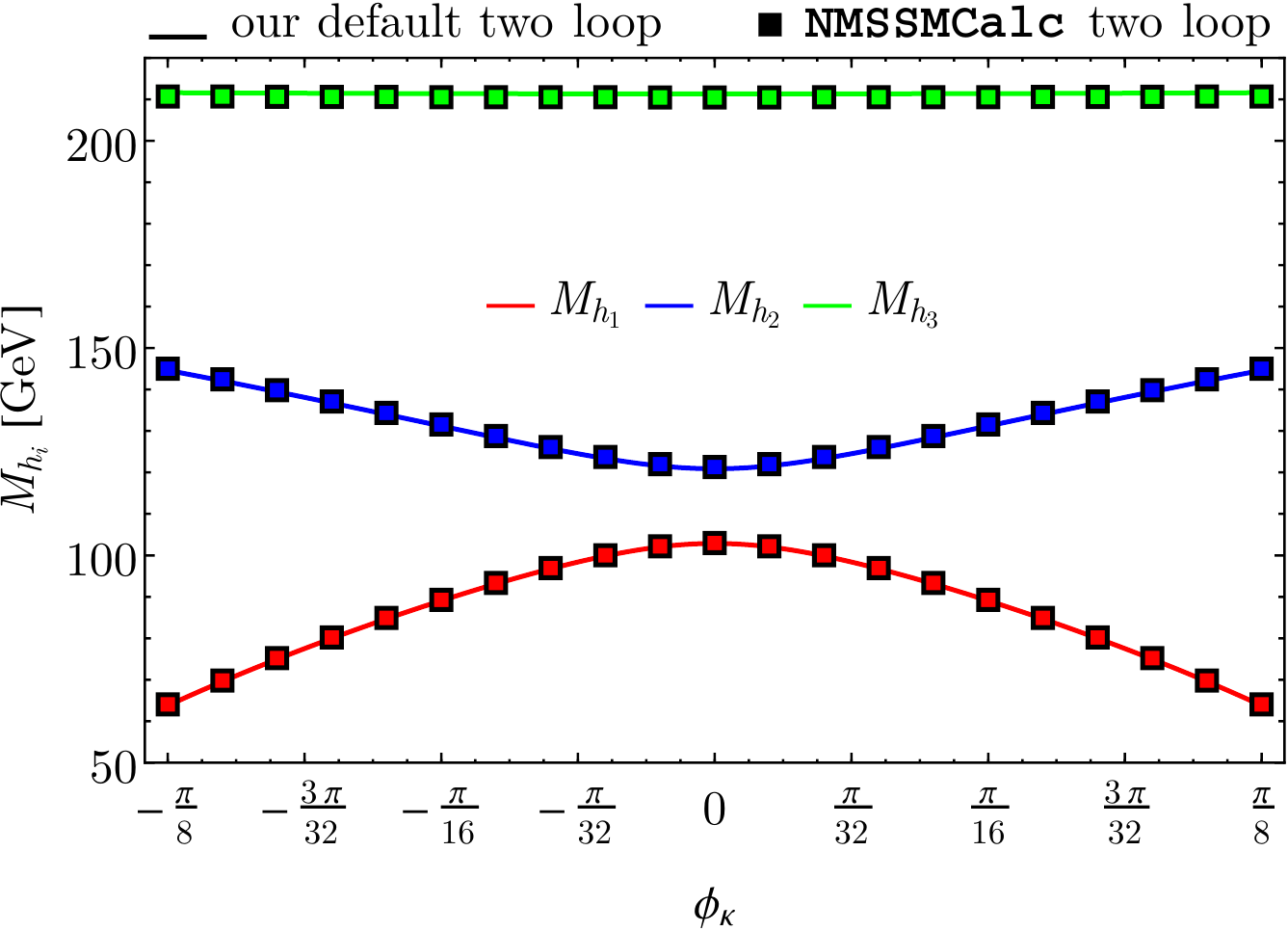}
  \caption{\label{fig:NMSSMmass1}  Masses of  the three  lighter Higgs
    fields   as  a   function  of   $\phi_\kappa$  for   the  scenario
    $\lambda=0.7$,        $|\kappa|=0.1$,        \mbox{$\tan\beta=2$},
    \mbox{$M_{H^{\pm}}=1170$\,GeV},                 $\mueff=500$\,GeV,
    $A_{\kappa}=-70$\,GeV,
    $m_{\tilde{Q}_3,\tilde{T},\tilde{B}}=0.5$\,TeV,
    $A_t=A_b=0.1$\,TeV,  \mbox{$2\,M_1=M_2=M_3/5=0.5$\,TeV}.  The  red
    color  depicts  the  mass  of the  mostly  \cp-odd,  singlet-like
    state~$h_1$; blue  is associated to the  essentially SM-like state
    $h_2$ and  green corresponds  to the mostly  \cp-even singlet-like
    state $h_3$.   We display  our ``default''  result for  the masses
    (solid curves) as well as the predictions of \NC\ (squares).}
  \end{center}
  \vspace{4ex}
\end{figure}

The   results    for   the   mass   prediction    are   presented   in
Fig.~\ref{fig:NMSSMmass1}.  At  vanishing $\phi_{\kappa}$ the  mass of
the SM-like state (in blue)  is somewhat low, $m_{h_2} \sim 120$\,GeV,
so that  this point in parameter  space has a very  marginal agreement
with   the  observed   characteristics  of   the  Higgs   state.   For
non-vanishing  $\phi_{\kappa}$, however,  a \cp-violating  mixing with
the  lighter  pseudoscalar  singlet  (in red)  develops:  this  effect
increases  the mass  of  the  light mostly  \cp-even  state $h_2$  but
affects its  otherwise SM-like  properties only  in a  subleading way.
Consequently,  we   recover  an  excellent  agreement   with  the  LHC
results---as  tested   by  \verb|HiggsSignals|  and   \HB---for  \EG{}
$\phi_{\kappa}\simeq-0.11$.   Additionally,   the  dominantly  \cp-odd
singlet $h_1$ then  has a mass close to $100$\,GeV.   As it acquires a
doublet \cp-even component via mixing,  it could explain the LEP local
excess  in $b\bar{b}$  final states~\cite{Barate:2003sz}.   The mostly
\cp-even singlet $h_3$ (in green), with mass at $\sim210$~GeV plays no
significant role.  The masses of the heavier doublet-like fields $h_4$
and $h_5$ are approximately constant and close to $M_{H^\pm}$.

\begin{figure}[t]
  \includegraphics[width=\linewidth]{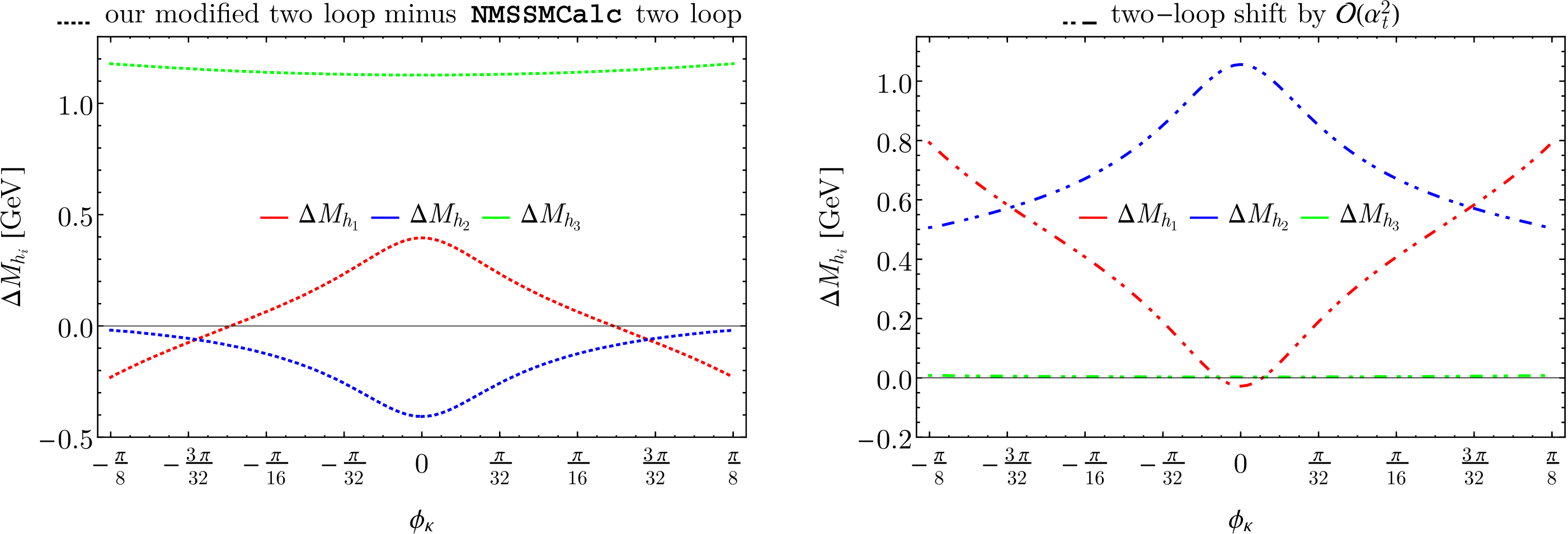}
  \caption{\label{fig:NMSSMmass12} Impact of two-loop contributions in
    the scenario of Fig.~\ref{fig:NMSSMmass1}.  On the left-hand side,
    \mbox{$\Delta{M_{h_i}} = M_{h_i}  - M_{h_i}^{\NCs}$} correspond to
    the mass-differences (for each of the three lightest Higgs states)
    between  the  predictions  of  our ``modified''  scheme  and  \NC:
    $\mathcal{O}{\left(\alpha_t\alpha_s\right)}$ should dominate these
    deviations. On the  right-hand side, $\Delta{M_{h_i}}$ corresponds
    to          the         mass-shifts          associated         to
    $\mathcal{O}{\left(\alpha_t^2\right)}$  contributions,  which  are
    calculated  in our  ``default'' scheme.  The color  of the  curves
    match the convention of Fig.~\ref{fig:NMSSMmass1}.}
\end{figure}

In Fig.~\ref{fig:NMSSMmass1}, we observe  a good agreement between our
results     (solid     lines),     computed    as     described     in
section~\ref{sec:theory},  and  the  predictions  of  \NC\  (squares),
although the corresponding masses are defined in different schemes and
at different orders.   For a more quantitative comparison,  we turn to
our ``modified''  scheme for the  mass calculation.  On  the left-hand
side of Fig.~\ref{fig:NMSSMmass12}, we  plot the deviation between the
corresponding  results  and the  predictions  of  \NC\ for  the  three
lightest  Higgs states.   We  checked that  the  one-loop results  are
virtually  identical, so  that the  differences between  \NC\ and  our
calculation are  entirely controlled by two-loop  effects.  We observe
typical deviations of order $0.5$--$1$\,GeV that should be interpreted
as   the    impact   of   $\mathcal{O}{\left(\alpha_t\alpha_s\right)}$
corrections beyond  the MSSM-approximation. As could  be expected, the
masses of  the mostly  singlet states  (red and  green lines)  tend to
exhibit  the  largest  effect,  though the  mass-predictions  for  the
SM-like   state    may   still    differ   by    $\sim0.5$\,GeV   (for
$\phi_{\kappa}\simeq0$).   The   plot  on   the  right-hand   side  of
Fig.~\ref{fig:NMSSMmass12}      depicts      the     magnitude      of
$\mathcal{O}{\left(\alpha_t^2\right)}$ effects, which is quantified in
our ``default'' scheme.  Here again,  the typical impact on the masses
is  of order  $1$\,GeV.  Expectedly,  the masses  of  the almost  pure
singlet states  (red curve  at $\phi_{\kappa}\simeq0$ or  green curve)
are   insensitive    to   the    corrections   implemented    in   the
MSSM-approximation.  The mass of the mostly \cp-odd singlet (red curve)
is only affected when the corresponding state acquires a non-vanishing
doublet component ($\phi_{\kappa}\neq0$).

\begin{figure}[t]
  \includegraphics[width=.5\linewidth]{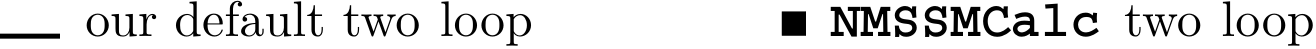}\\
  \includegraphics[width=\linewidth]{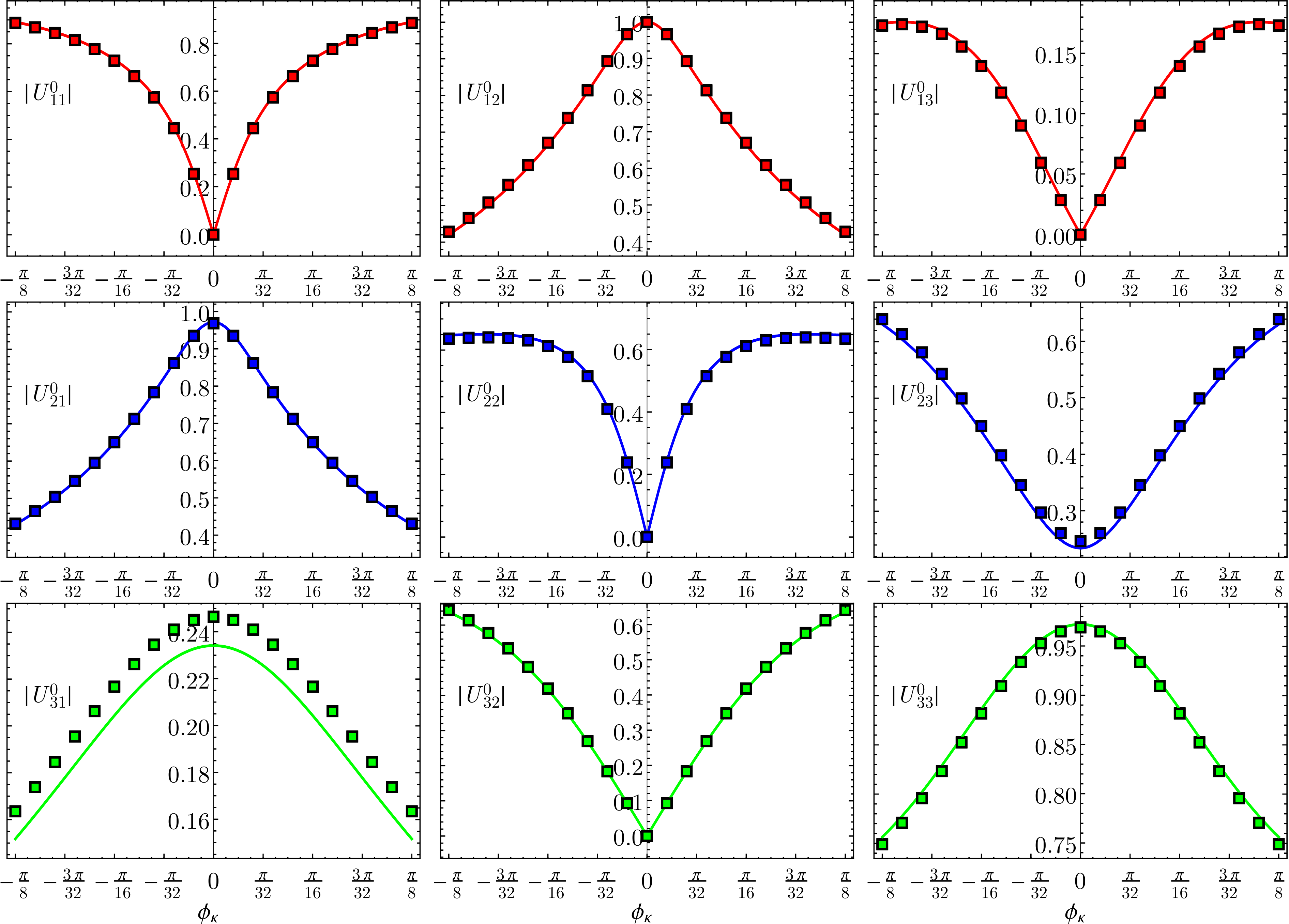}
  \caption{\label{fig:NMSSMU01}The matrix-elements $|U^0_{ij}|$ in our
    calculation (full curves)  and in \NC\ (squares)  for the scenario
    of Fig.~\ref{fig:NMSSMmass1}.}
\end{figure}

In  Fig.~\ref{fig:NMSSMU01},  we  compare  the  $\mathbf{U}^0$  matrix
elements that are  delivered by \NC\ (squares) with  ours (solid line;
\NC\ does  not provide $\mathbf{Z}^{\mbox{\tiny mix}}$).   The results
show a satisfactory agreement also at this level.

\medskip

Subsequently,  we  present  our  results  in  another  region  of  the
parameter   space:   $\lambda=0.2$,  $|\kappa|=0.6$,   $\tan\beta=25$,
$m_{H^{\pm}}=1$\,TeV,  $\mueff=200$\,GeV, $A_{\kappa}=-750$\,GeV,  the
gaugino soft  masses as well as  the soft masses for  the sfermions of
first  and second  generations are  chosen  as before;  for the  third
generation, the soft sfermion mass is set to $1.1$\,TeV; the trilinear
soft terms are  set to $-2$\,TeV. With this choice  of parameters, the
singlet-like  \cp-even  state  and  the  heavy  \cp-even  and  \cp-odd
doublet-like  states  receive  comparable   masses  of  the  order  of
$1$\,TeV.  This  results in  a sizable  mixing for  the corresponding
fields $h_2$,  $h_3$ and  $h_4$, which includes  both singlet--doublet
admixture    as    well    as   \cp-violation    (for    non-vanishing
$\phi_{\kappa}$).   The  SM-like  Higgs  state has  a  mass  close  to
$\sim124$\,GeV on the whole range of $\phi_{\kappa}$, which leads to a
good  agreement with  the Higgs  properties  measured at  the LHC  (as
tested with \verb|HiggsSignals|).  The heaviest state $h_5$ has a mass
of $\sim1.2$\,TeV, which we will not comment further below.

\begin{figure}[t]
  \includegraphics[width=\linewidth]{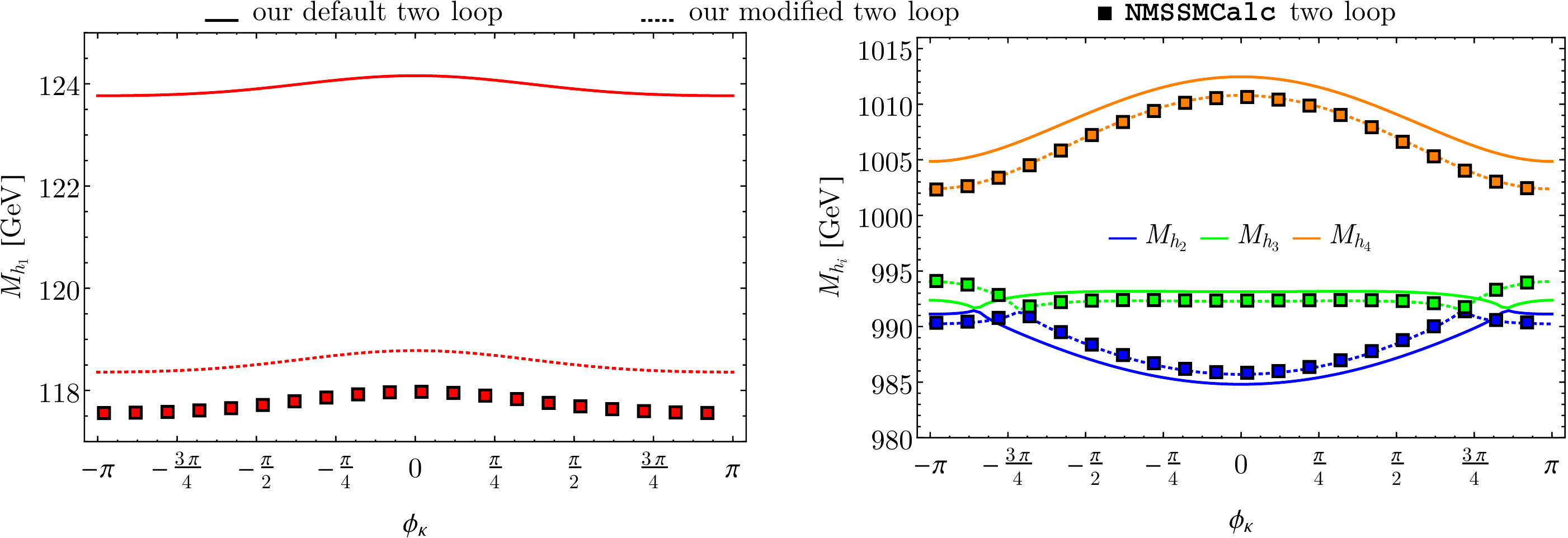}
  \caption{\label{fig:NMSSMmass21} Mass  predictions as a  function of
    $\phi_\kappa$ for  the lightest, mostly SM-like  Higgs state $h_1$
    on the left-hand side, and  the states $h_2$ (blue), $h_3$ (green)
    and $h_4$  (orange) on  the right-hand  side.  The  latter involve
    essentially the \cp-even singlet and  the two \cp-even and \cp-odd
    heavy-doublet degrees of freedom  that mix substantially. The full
    curves correspond to our default  result; the squares are obtained
    with \NC;  the dotted  line represents  our modified  result.  The
    parameters are  chosen as follows:  $\lambda=0.2$, $|\kappa|=0.6$,
    $\tan\beta=25$,      $m_{H^{\pm}}=1$\,TeV,      $\mueff=200$\,GeV,
    $A_{\kappa}=-750$\,GeV,
    $m_{\tilde{Q}_3,\tilde{T},\tilde{B}}=1.1$\,TeV, $A_t=A_b=-2$\,TeV,
    $2\,M_1=M_2=M_3/5=0.5$\,TeV.}
\end{figure}

On  the  left-hand side  of  Fig.~\ref{fig:NMSSMmass21},  we show  our
prediction for  the mass of  the lightest (SM-like) Higgs  state (full
curve).  The mass  delivered by \NC\ is represented by  the squares at
about~$118$\,GeV,  which  is  substantially   smaller  than  ours  (by
$\approx6$\,GeV).  If we  mimic the settings of  \NC\ (our ``modified
result'', dotted curve), this discrepancy is considerably reduced.  In
fact,  the difference  between our  full result  and \NC's  is largely
driven   by    the   $\mathcal{O}{\left(\alpha_t^2\right)}$   two-loop
contributions,  missing in  \NC.   Again, both  results are  virtually
identical at the one-loop order.

\begin{figure}[t]
  \includegraphics[width=\linewidth]{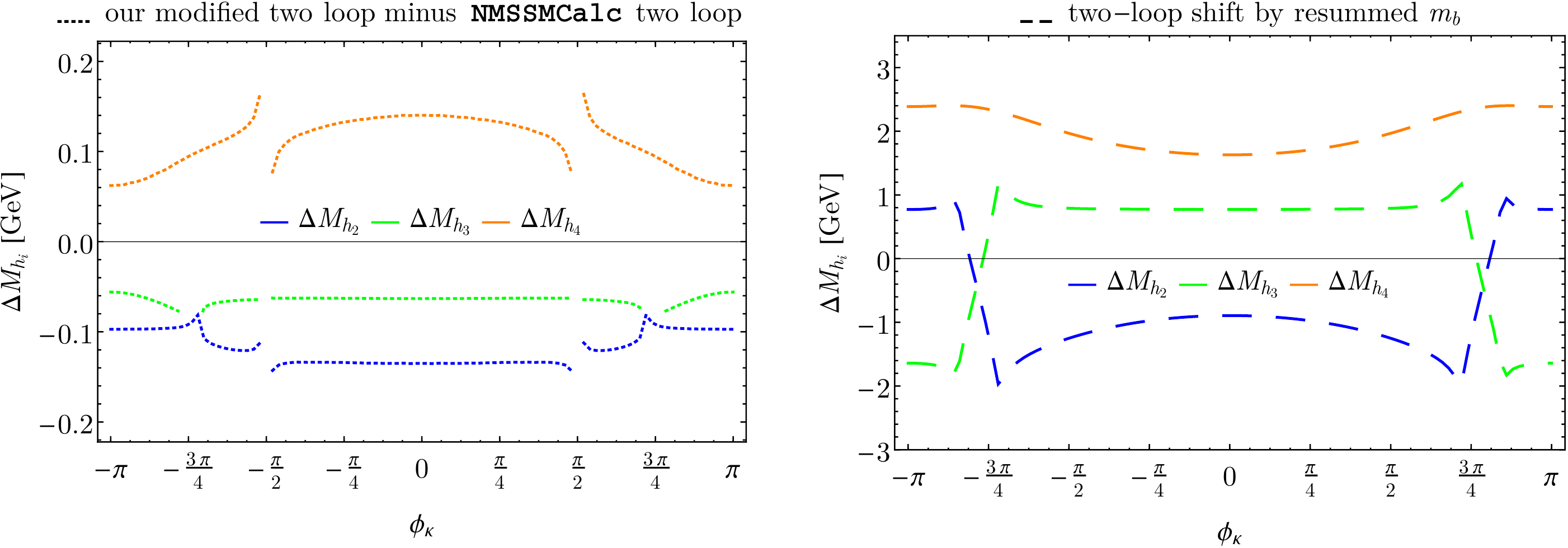}
  \caption{\label{fig:NMSSMmass2} Impact  of the  higher-order effects
    in the mass-predictions for $h_2$, $h_3$ and $h_4$ in the scenario
    of Fig.~\ref{fig:NMSSMmass21}. On the  left-hand side, we show the
    deviation in mass between \NC\ and our modified scheme (controlled
    by $\mathcal{O}{\left(\alpha_t\alpha_s\right)}$ corrections beyond
    the  MSSM-approximation).    The  plot  on  the   right-hand  side
    illustrates the impact of  large-$\tan\beta$ effects in $Y_b$: the
    considered mass-difference  is that induced in  our default scheme
    by  the $\Delta_b$  term. The  colors follow  the conventions  of
    Fig.~\ref{fig:NMSSMmass21}.           The          discontinuities
    at~$\phi_\kappa=\pm   \frac{\pi}{2}$  originate   from  the   mass
    calculation in \NC.}
\end{figure}

On the right-hand  side of Fig.~\ref{fig:NMSSMmass21}, we  turn to the
heavier states $h_2$,  $h_3$ and $h_4$ of this  scenario.  Our default
results  (full   curves)  are  compatible  with   the  predictions  of
\NC\  (squares). The  discrepancies are  of order  $1$--$3$\,GeV only,
which should be considered from  both the perspective of the different
renormalization scheme of the electric  coupling $e$ and the different
two-loop contributions.   Actually, the mass predictions  match almost
exactly when comparing \NC\ with  our modified scheme (dotted curves).
The  corresponding  deviations are  shown  on  the left-hand  side  of
Fig.~\ref{fig:NMSSMmass2} and fall in the range of $100$\,MeV. In this
precise  case,  the  difference  between   our  results  and  \NC\  is
essentially driven by the  resummation of large-$\tan\beta$ effects in
the   $b$-quark  Yukawa   coupling.    On  the   right-hand  side   of
Fig.~\ref{fig:NMSSMmass2}, we  quantify the associated  mass-shift and
find an impact of a few~GeV.

\begin{figure}[b]
  \begin{center}
  \includegraphics[width=.8\linewidth]{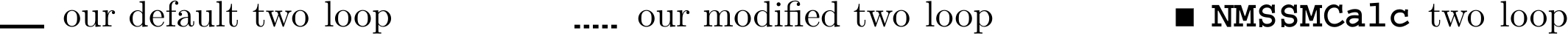}\\
  \includegraphics[width=\linewidth]{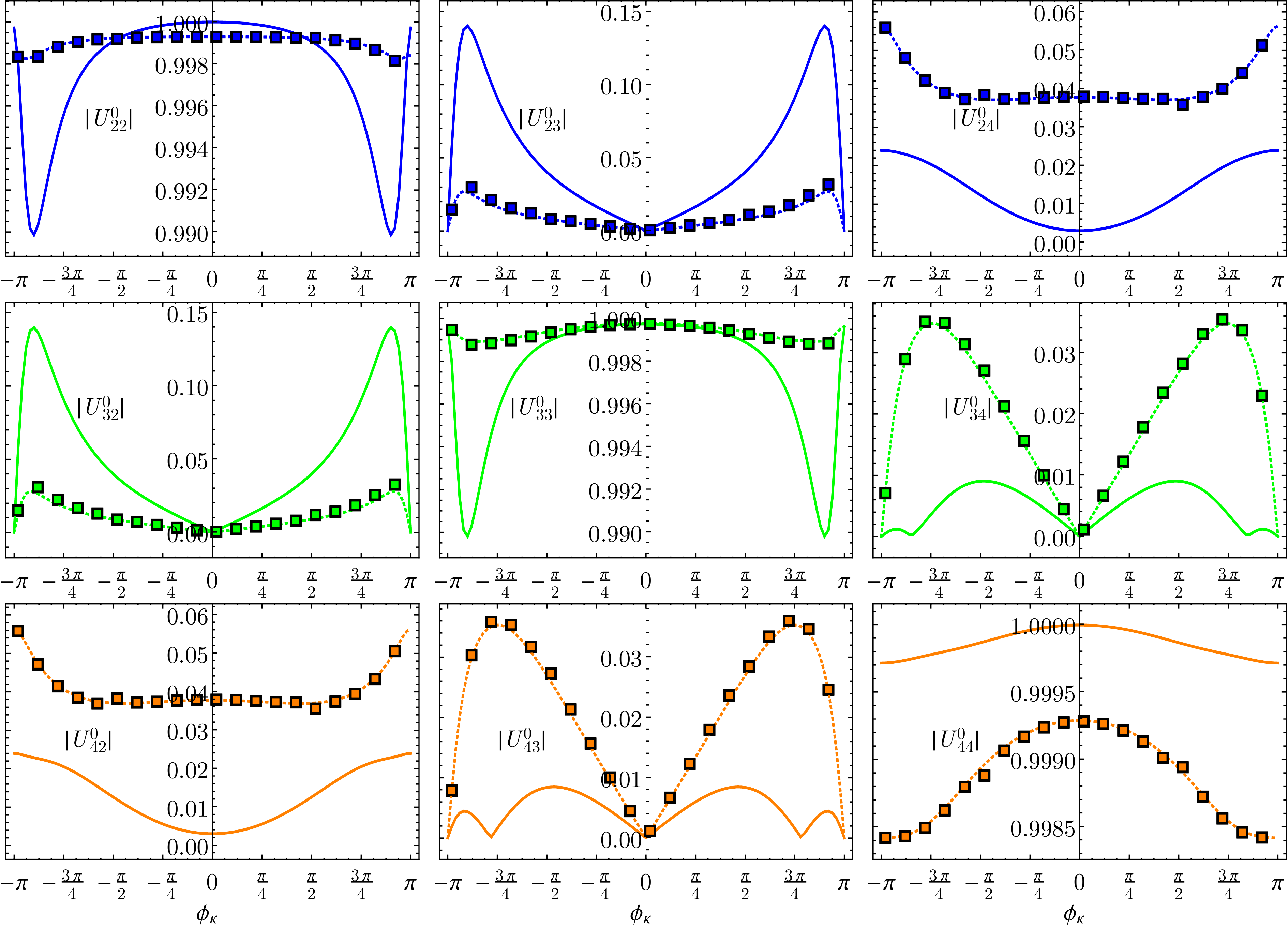}
  \end{center}\vspace{-2ex}
  \caption{\label{fig:NMSSMU02}     The    matrix-elements     $\lvert
    U^0_{ij}\rvert$   in   our    calculation   (solid   curves),   in
    \NC\  (squares), and  in our  modified calculation  closer to  the
    options    of    \NC\    (dotted)     for    the    scenario    of
    Fig.~\ref{fig:NMSSMmass21}.}
  \vspace{2ex}
\end{figure}

Finally,  we turn  to the  $\mathbf{U}^0$ matrix  elements for  $h_2$,
$h_3$ and $h_4$ in Fig.~\ref{fig:NMSSMU02}.  There, we observe sizable
deviations   between   our   default   result   (solid   curves)   and
\NC\ (squares), which,  however, have no deep-reason to  agree in view
of the diverging options. If we  keep in mind that the main difference
between   our   full   scheme   and  \NC\   is   controlled   by   the
large-$\tan\beta$ corrections to $Y_b$ in this precise scenario, it is
not surprising  to observe large  shifts, as mixing angles  are indeed
very sensitive to small deviations  in the mass-matrix for states that
are  very close  in mass.   These differences  largely vanish  when we
identify the output of \NC\  with our modified results (dotted lines),
which is better equipped for comparisons with this code.

In summary, the results of our mass calculation are largely compatible
with   the  predictions   of   \NC.   Deviations   with  a   magnitude
of~$\mathcal{O}{\left(\text{GeV}\right)}$   are   indeed  within   the
expected range  if we allow  for the different  renormalization scheme
of~$e$ and higher-order contributions.   Such discrepancies tend to be
reduced sizable when  we modify our routines to  adopt the assumptions
of         \NC.          The        impact         of         two-loop
$\mathcal{O}{\left(\alpha_t\alpha_s\right)}$ contributions  beyond the
MSSM,   two-loop    $\mathcal{O}{\left(\alpha_t^2\right)}$   and   the
resummation of  large-$\tan\beta$ effects in the  Yukawa couplings can
be clearly identified from this comparison.  As a final remark, let us
emphasize  that the  uncertainty  on the  Higgs-mass calculation  from
unknown  higher-order contributions  and parametric  errors may  reach
several~GeV~\cite{Degrassi:2002fi,Staub:2015aea,Drechsel:2016htw}.

\subsection[The matrix $\mathbf{Z}^{\mbox{\tiny mix}}$ and the $h_i\to\tau^+\tau^-$ decays]{\boldmath The matrix $\mathbf{Z}^{\mbox{\tiny mix}}$ and the $h_i\to\tau^+\tau^-$ decays}

The  matrix  $\mathbf{Z}^{\mbox{\tiny  mix}}$  is  not  an  observable
quantity in  itself.  It is a  renormalization-scheme dependent object
relating  the  tree-level mass  states  of  the  Higgs sector  to  the
physical   Higgs    fields.    For   on-shell    renormalized   fields
$\mathbf{Z}^{\mbox{\tiny   mix}}$   is    trivial.    In   any   other
renormalization  scheme,  however, it  is  mandatory  to include  this
transition to the physical fields for a proper description of external
legs in Feynman diagrams at higher orders.

A  remarkable aspect  of $\mathbf{Z}^{\mbox{\tiny  mix}}$ is  that the
eigenvectors that it  contains do not preserve  unitarity with respect
to  the tree-level  fields.  Instead,  they satisfy  the normalization
condition given in  Eq.~\eqref{physnorm}.  This is a  feature that the
approximations $\mathbf{U}^0$ and $\mathbf{U}^m$ are unable to capture
(by construction).   In a first step,  we will show that  the norms in
$\mathbf{Z}^{\mbox{\tiny mix}}$ can  differ from $1$ by  a few percent
in  the scheme  that  we have  described in  section~\ref{sec:theory}.
Beyond   the  normalization   of   the   fields,  $\mathbf{U}^0$   and
$\mathbf{U}^m$  also differ  from  $\mathbf{Z}^{\mbox{\tiny mix}}$  in
that  they diagonalize  the mass-matrix  away  from the  poles of  the
propagator.

However,  as  we wrote  above,  $\mathbf{Z}^{\mbox{\tiny  mix}}$ is  a
scheme-dependent object and  we should not pay  excessive attention to
its  actual  structure.  In  order  to  characterize  its role  in  an
observable quantity, we will  consider the $h_i\to\tau^+\tau^-$ decays
at the one-loop  level.  We have chosen this particular  channel as it
is  one of  the main  fermionic  Higgs decays  and proves  technically
simple  to   implement  in  a  predictive   way.   Moreover,  one-loop
corrections are of purely electroweak nature---QCD contributions occur
only at  three-loop order  and beyond---so that  radiative corrections
are   expected   to   be   moderate.   This   allows   for   a   clean
appreciation---free  of  large   higher-order  uncertainties---of  the
impact        of         the        wave-function        normalization
matrix~$\mathbf{Z}^{\mbox{\tiny           mix}}$.            Radiative
corrections\footnote{Details on  the calculation of the  decays at the
  one-loop  level will  be presented  in a  future publication.}   are
computed with our model file,  except for the QED contributions, which
are     included     according     to     the     prescriptions     of
Refs.~\cite{Braaten:1980yq,           Drees:1990dq}.            There,
$\mathbf{Z}^{\mbox{\tiny mix}}$ intervenes in  the decay amplitudes of
the physical fields  according to Eq.~\eqref{eq:RelationPhysAmplitude}
(we dismiss the superscript 'phys'  throughout this section).  We will
show that  the substitution of $\mathbf{Z}^{\mbox{\tiny  mix}}$ by the
approximations $\mathbf{U}^0$  and $\mathbf{U}^m$ may lead  to sizable
deviations in certain  regions of the the NMSSM  parameter space. This
result   confirms   the   outcome    of   similar   studies   in   the
MSSM~\cite{Frank:2006yh}.

\begin{figure}[t]
  \vspace{1ex}
  \includegraphics[width=\linewidth]{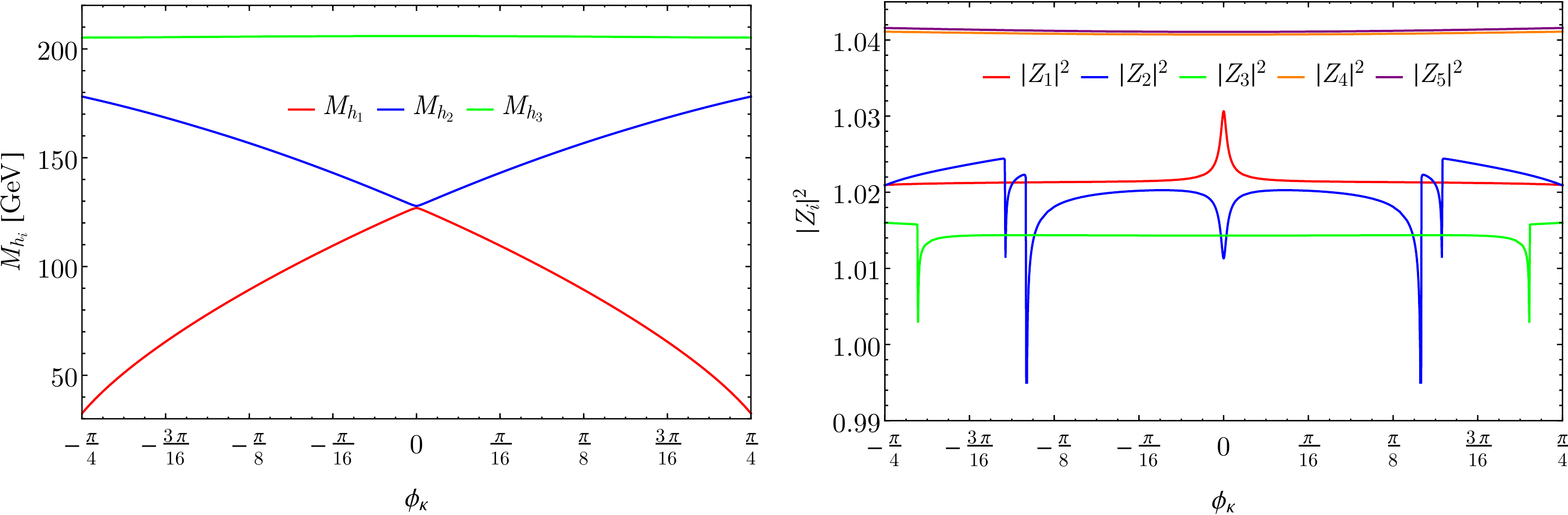}
  \caption{\label{fig:NMSSMZ} On the left-hand side, the masses of the
    three  lighter  Higgs  states  are   depicted  as  a  function  of
    $\phi_\kappa$, including all  available two-loop contributions. On
    the right-hand  side, the  squared norms  $\left|Z_i\right|^2$ (as
    defined in Eq.~\eqref{eq:norm}) of  the five eigenvectors defining
    $\mathbf{Z}^{\mbox{\tiny  mix}}$  are   shown.   The  scenario  is
    characterized  by  $\lambda=0.7$,  $|\kappa|=0.1$,  $\tan\beta=2$,
    $m_{H^{\pm}}=1.2$\,TeV, $\mueff=500$\,GeV, $A_{\kappa}=-100$\,GeV,
    $m_{\tilde{Q}_3,\tilde{T},\tilde{B}}=0.5$\,TeV,
    $A_t=A_b=0.1$\,TeV, $2\,M_1=M_2=M_3/5=0.5$\,TeV.}
\end{figure}

\begin{figure}[t!]
  \begin{center}\vspace{-2ex}
  \includegraphics[width=.9\linewidth]{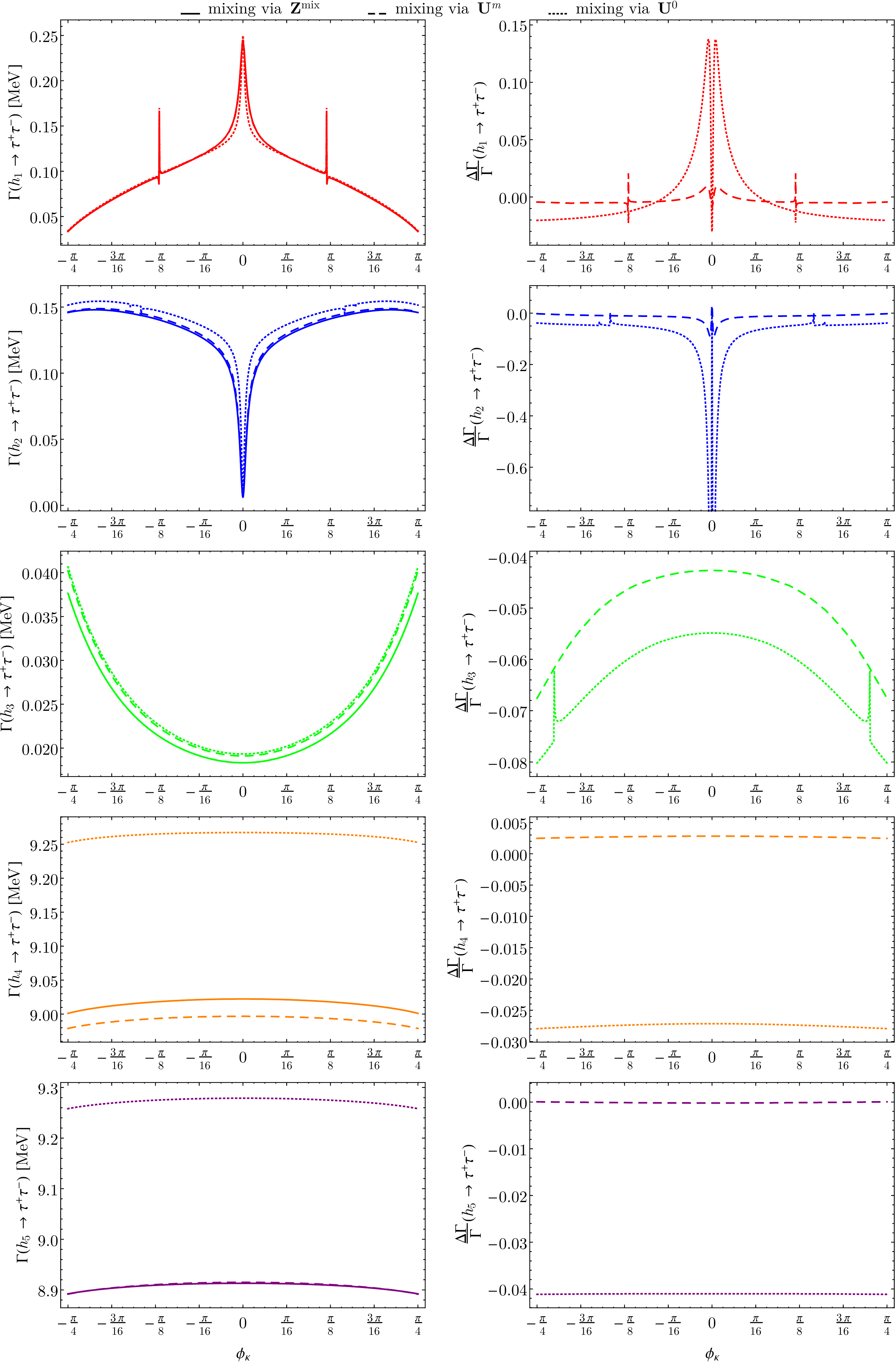}
  \end{center}\vspace{-3ex}
  \caption{\label{fig:NMSSMGam} In the left  column, we show the decay
    widths   $\Gamma(h_i\to\tau^+\tau^-)$    in   the    scenario   of
    Fig.~\ref{fig:NMSSMZ}  for the  five  neutral  Higgs states.   The
    widths are computed  at the one-loop level, and the  mixing of the
    external,  physical Higgs  fields  is expressed  in  terms of  the
    matrix   \Zmix\  (solid),   or   approximated   by  the   matrices
    $\mathbf{U}^m$ (dashed) or $\mathbf{U}^0$  (dotted).  In the right
    column,    the   differences    $\Delta{\Gamma}    =   \Gamma    -
    \Gamma_{\textrm{appr}}$   between   the   widths   obtained   with
    \Zmix\ and its approximate  treatments $\mathbf{U}^m$ (dashed) and
    $\mathbf{U}^0$  (dotted) are  depicted,  normalized  to the  width
    $\Gamma$ obtained with \Zmix.}
\end{figure}

We turn to the following NMSSM  input: the parameters are chosen as in
Fig.~\ref{fig:NMSSMmass1},   except  for   $M_{H^{\pm}}=1.2$\,TeV  and
$A_{\kappa}=-100$\,GeV.   We  have plotted  the  masses  of the  three
lighter Higgs fields as a function of $\phi_\kappa$ in the plot on the
left-hand    side    of    Fig.~\ref{fig:NMSSMZ}.     For    vanishing
$\phi_{\kappa}$  the lightest  Higgs  state $h_1$  is  SM-like and  we
checked  with  \verb|HiggsBounds|  and \verb|HiggsSignals|  that  this
point  is  consistent  with  the experimental  data.   The  dominantly
\cp-odd, singlet-like  state $h_2$ is  only slightly heavier  than the
state $h_1$  in this case.   For increasing values  of $\phi_{\kappa}$
the mixing of  the states $h_1$ and  $h_2$ tends to lower  the mass of
the  SM-like  state  $h_1$,  which eventually  becomes  too  light  to
accommodate   the   experimental   data.   The   dominantly   \cp-even,
singlet-like state  $h_3$ has a  near constant mass  of $\sim210$\,GeV
for all depicted  values of $\phi_\kappa$.  The  two heavier, \cp-even
and \cp-odd doublet-like states have masses close to $\sim 1.2$\,TeV.

The  results  for  the  squared  norms  $\lvert  Z_i\rvert^2$  of  the
eigenvectors---see Eq.~\eqref{eq:norm}---in this scenario are shown in
the plot on the right-hand  side of Fig.~\ref{fig:NMSSMZ}.  We observe
a departure from  the value~$1$---which would correspond  to a unitary
transition,  as  modeled  by  the  approximations  $\mathbf{U}^0$  and
$\mathbf{U}^m$---by   a   few   percent.    The   local   extrema   at
$\phi_{\kappa}\simeq0$  for  $\lvert   Z_1\rvert^2$  (red  curve)  and
$\lvert  Z_2\rvert^2$  (blue  curve)  are  associated  to  the  sudden
disappearance of the mixing between  the light \cp-odd singlet and the
SM-like  states  at $\phi_\kappa  =  0$  (\cp-conserving limit).   The
discontinuities  of  $\lvert  Z_2\rvert^2$  and  $\lvert  Z_3\rvert^2$
(green curve)  at $\phi_{\kappa}\simeq\pm0.5$  and~$\pm0.7$ correspond
to the crossing  of decay thresholds ($h_2\to  W^+W^-$, $h_2\to 2\,Z$,
$h_{3}\to  h_1\,h_2$).    These  ``spikes''  are  associated   to  the
singularities of the first derivatives  of the loop functions involved
in the  determination of  \Zmix---the apparent  singularities actually
come with a finite height due to  the imaginary parts of the poles.  A
proper description of  these threshold regions would  require that the
interactions among the daughter particles (of the decays at threshold)
are  properly  taken  into  account,   which  would  result  in  \EG{}
interactions between the Higgs state  and bound-states or $s$-waves of
the daughter  particles. This,  however goes beyond  the scope  of the
present work.

We   now   turn  to   the   decay   widths  $\Gamma{(h_i   \rightarrow
  \tau^+\tau^-)}$ in the scenario of Fig.~\ref{fig:NMSSMZ}. The widths
are displayed  in the  left column  of Fig.~\ref{fig:NMSSMGam}  in the
exhaustive  description of  the  Higgs external  leg (\IE{}  employing
\Zmix;  solid curves),  in  the  $\mathbf{U}^m$ approximation  (dashed
lines) and in the $\mathbf{U}^0$ approximation (dotted lines), for the
five Higgs  mass-eigenstates.  We observe  a sharp variation  close to
$\phi_{\kappa}=0$ for the decays of $h_1$  and $h_2$, both in the full
and approximate  descriptions.  It  is associated  to the  mixing that
develops between the  SM-like state $h_1$ and  the dominantly \cp-odd,
singlet-like state  $h_2$: this effect  transfers part of  the doublet
component  of $h_1$  to $h_2$,  so that  the second  state acquires  a
non-vanishing coupling to SM fermions at the expense of the first. The
sum of  the decay widths  for both these states  remains approximately
constant   in   the   vicinity  of   $\phi_{\kappa}=0$.    The   width
$\Gamma(h_3\to\tau^+\tau^-)$  appears  to  be an  order  of  magnitude
smaller than the  corresponding widths for $h_1$ and  $h_2$, an effect
that is associated to the  dominantly \cp-even, singlet-like nature of
$h_3$.   Still,  $\Gamma(h_3\to\tau^+\tau^-)$  nearly doubles  in  the
considered interval  of $\phi_{\kappa}$,  while the  mass of  $h_3$ is
fairly stable: we can understand this fact in terms of the acquisition
of a  larger doublet  component, which is  channeled by  the increased
proximity of the  masses of $h_2$ and $h_3$.  The  widths of $h_4$ and
$h_5$ are essentially constant with  only small relative changes.  The
general  $\phi_\kappa$-dependency of  the  approximated  and the  full
results are  very similar.  Yet,  a systematic shift can  be observed,
especially in the case of  $\mathbf{U}^0$. This is consistent with the
findings    of   similar    studies    in   the    context   of    the
MSSM~\cite{Frank:2006yh}.

On  the  right-hand  side  of  Fig.~\ref{fig:NMSSMGam},  we  show  the
difference   between   the   full   and   the   approximate   results,
$\Delta{\Gamma} = \Gamma -  \Gamma_{\textrm{appr}}$, normalized to the
more accurate one obtained with \Zmix.  When \Zmix\ is approximated by
$\mathbf{U}^0$ (dotted lines), the typical discrepancy averages $4\%$,
although the deviation reaches beyond $10\%$ in the case of the decays
of  the two  lightest Higgs  states in  the vicinity  of, but  not at,
$\phi_{\kappa}\simeq0$.   We  stress  that   this  interval  close  to
$\phi_{\kappa}=0$  corresponds  to   the  phenomenologically  relevant
region from  the perspective  of the  measured Higgs  properties.  The
approximation  of \Zmix\  by  $\mathbf{U}^m$ tends  to provide  better
estimates of the full result, though deviations reach up to $\sim7\%$.
For both approximations the largest  deviations from the more complete
result employing  \Zmix\ are  intimately related  to the  proximity in
mass of  the SM-like and  dominantly \cp-odd, singlet-like  states: as
the  approximations capture  the dependence  on the  external momentum
either partially ($\mathbf{U}^m$) or  not at all ($\mathbf{U}^0$), the
gap between the diagonal elements  of the Higgs mass-matrix, hence the
mixing between the two states,  is not quantified properly. While this
precise  configuration might  appear  somewhat anecdotal,  we wish  to
point  out   the  popularity  of   NMSSM  scenarios  with   a  sizable
singlet--doublet   mixing.    Dismissing   this  extreme   case,   the
approximations of \Zmix\ by $\mathbf{U}^0$,  and to a lesser extent by
$\mathbf{U}^m$, still generate errors of the order of a few percent at
the  level of  the decay  widths.   In view  of the  precision of  the
measurements   achievable    at   the    LHC~\cite{CMS-HL,   ATLAS-HL,
  Khachatryan:2016vau,    Zenz:2016lia,    Slawinska:2016zeh},    such
discrepancies may  appear of secondary  importance.  In the  long run,
however, if the  Higgs couplings are studied more closely,  \EG{} at a
linear   collider~\cite{Baer:2013cma,  Fujii:2015jha,   Fujii:2017ekh,
  Moortgat-Picka:2015yla}, one would have to try and keep such sources
of error to a minimum.
\clearpage

\section{Conclusions}
\label{sec:conslusion}

In  this paper,  we have  discussed the  renormalization of  the NMSSM
Higgs sector, including complex parameters. Radiative contributions to
the Higgs self-energies have been  included up to the leading two-loop
MSSM-like                          effects                          of
$\mathcal{O}{\left(\alpha_t\alpha_s+\alpha_t^2\right)}$.   Beyond  the
calculation  of  on-shell  Higgs  masses   in  this  scheme,  we  were
interested      in     determining      the     transition      matrix
$\mathbf{Z}^{\mbox{\tiny  mix}}$  between  the  mass-  and  tree-level
states.   The latter  plays  an essential  role  in the  proper
description  of  external Higgs  legs  in  physical processes  at  the
radiative level.

Our  predictions  for the  Higgs  masses  have  been compared  to  the
calculations  of existing  tools in  several NMSSM  scenarios. In  the
MSSM-limit of the model, we have recovered an excellent agreement with
\verb|FeynHiggs|. For  non-vanishing $\lambda$ and $\kappa$,  we first
compared our  Higgs-mass prediction  with the  findings of  a previous
extension  of  \verb|FeynHiggs| to  the  NMSSM  in  the case  of  real
parameters, and  found nearly  identical results. Second,  we compared
our   calculation   in   the   case   of   complex   parameters   with
\verb|NMSSMCalc|  and  found values  of  the  Higgs masses  which  are
compatible, although small differences emerge as a result of different
processing of the two-loop pieces, both for low and high $\tan\beta$.

Finally,  we   investigated  the  impact  of   the  transition  matrix
$\mathbf{Z}^{\mbox{\tiny mix}}$ on the $h_i\to\tau^+\tau^-$ width in a
scenario with low  $\tan\beta$ and large $\lambda$,  where the SM-like
Higgs state  may have a  sizable mixing  with the \cp-odd  singlet. We
compared the full one-loop  calculation of the width---\IE{} including
$\mathbf{Z}^{\mbox{\tiny  mix}}$---with   the  popular  approximations
$\mathbf{U}^0$  and $\mathbf{U}^m$---which  are determined  for fixed,
unphysical momenta.  We found typical deviations at the percent level,
although   larger   effects   can   develop   in   the   presence   of
almost-degenerate   states,    especially   in    the   $\mathbf{U}^0$
approximation. Such precision effects will matter when the measurement
of fermionic Higgs couplings reaches comparable accuracy.

In its  current form,  our mass-computing tool  is contained  within a
\texttt{Mathematica}  package.   In  time,   our  routines  should  be
incorporated in an extension of \verb|FeynHiggs| to the NMSSM.

\section*{Acknowledgments}

We thank  S.~Heinemeyer for  helpful comments  on the  manuscript. The
work  of  F.~Domingo  is  supported   in  part  by  CICYT  (Grant  FPA
2013-40715-P),   in  part   by  the   MEINCOP  Spain   under  contract
FPA2016-78022-P,  and in  part  by the  Spanish  ``Agencia Estatal  de
Investigación''  (AEI)  and  the  EU  ``Fondo  Europeo  de  Desarrollo
Regional'' (FEDER)  through the  project FPA2016-78645-P. The  work of
P.~Drechsel and  S.~Paßehr is supported by  the Collaborative Research
Center  SFB676  of  the  DFG,   ``Particles,  Strings  and  the  early
Universe.''

\bibliographystyle{h-physrev}     
\bibliography{literature} 

\end{document}